\documentclass[aps,amssymb,amsmath,amsfonts,superscriptaddress,twocolumn,a4paper,accepted=2020-11-28]{quantumarticle}
\pdfoutput=1
\usepackage[utf8]{inputenc} 
\usepackage{amsmath} 
\usepackage{graphicx} 
\usepackage[usenames]{color} 
\usepackage{amssymb} 
\usepackage[toc,page,titletoc]{appendix}
\usepackage{bbm} 
\usepackage[hidelinks]{hyperref} 
\usepackage{minibox}
\usepackage{slashed}

\newcommand{\beq}{\begin{equation}}
\newcommand{\eeq}{\end{equation}}
\def\ba{\begin{eqnarray}}
\def\ea{\end{eqnarray}}

\def\barray{\begin{eqnarray}}
\def\earray{\end{eqnarray}}
\def\beq{\begin{equation}}
\def\eeq{\end{equation}}

\newcommand{\ket}{|\psi\ra}

\def\ra{{\rangle}}

\def\P{{|\mathbb{P}_n\ra}}

\def\Zmath{{\mathbb{Z}}}

\newcommand{\ie}{\textit{i.e.} }
\newcommand{\eg}{\textit{e.g.} }

\begin{document}
	
\setlength\parskip{5mm}

\title{The Prime state and its quantum relatives}

\author{Diego García-Martín}
\affiliation{Barcelona Supercomputing Center (BSC), Barcelona, Spain.}
\affiliation{Dept. F\'{i}sica Qu\`{a}ntica i Astrof\'{i}sica, Universitat de 
Barcelona, Barcelona, Spain.}
\affiliation{Instituto de F\'{i}sica Teórica, UAM-CSIC, Madrid, Spain.}
\author{Eduard Ribas}
\affiliation{Dept. F\'{i}sica Qu\`{a}ntica i Astrof\'{i}sica, Universitat de Barcelona, Barcelona, Spain.}
\author{Stefano Carrazza}
\affiliation{TIF  Lab,  Dipartimento  di  Fisica,  Universit\`a  degli  Studi  di  Milano  and  INFN  Milan, Milan, Italy.}
\affiliation{Quantum Research Centre, Technology Innovation Institute, Abu Dhabi, UAE.}
\author{Jos\'{e} I. Latorre}
\affiliation{Quantum Research Centre, Technology Innovation Institute, Abu Dhabi, UAE.}
\affiliation{Dept. F\'{i}sica Qu\`{a}ntica i Astrof\'{i}sica, Universitat de Barcelona, Barcelona, Spain.}
\affiliation{Center for Quantum Technologies, National University of Singapore, Singapore.}
\author{Germán Sierra}
\affiliation{Instituto de F\'{i}sica Teórica, UAM-CSIC, Madrid, Spain.}

\begin{abstract}
	
The Prime state of $n$ qubits, $\P$, is  defined as the uniform  superposition of all the computational-basis states corresponding to prime numbers  smaller than $2^n$.
This state encodes, quantum mechanically, arithmetic properties
of the primes. We first show that the  Quantum Fourier Transform of the Prime state provides a direct access to Chebyshev-like  biases in the distribution of prime numbers.  We next study the entanglement entropy 
of $\P$ up to $n=30$ qubits,  and find a relation between its scaling and the  Shannon entropy of the 
density of square-free integers.  This relation also holds   when the Prime state 
is constructed using a qudit basis, showing that this  property is  intrinsic to the distribution of primes.
The same feature is found when
considering states built from the superposition of primes in arithmetic progressions. 
Finally, we explore the properties of other number-theoretical quantum states, such as those defined from
odd composite numbers, square-free integers and starry primes.
For this study, we have developed an open-source library that diagonalizes matrices using floats of arbitrary precision.

\end{abstract}

\maketitle

\section{Introduction}

\noindent 
Mathematical sequences of integers are key to Number Theory and other areas of Mathematics. Properties of such sequences can be studied on a quantum computer by creating pure states consisting of  superpositions of computational-basis vectors that encode the numbers appearing in the sequence, \eg in binary format using qubits. We call these states \textit{number-theoretical quantum states}. This novel approach to sequences using the tools of Quantum Information Theory was introduced in Ref. \cite{prime}, where it was shown that a uniform superposition of prime numbers, \ie the Prime state, can be  created efficiently on a quantum computer by using Grover's search algorithm \cite{Grover} with a quantum oracle that codifies a primality test.

An example of the potential borne by the use of quantum methods to address problems related to prime numbers is provided by the 
possible numerical verification of the Riemann Hypothesis \cite{Riemann} on a quantum computer beyond what
can be achieved using  classical methods \cite{pi(x)}. A Prime state with a minimum of 90 logical qubits would be needed for that.
This is achieved by estimating, with a
quantum counting algorithm \cite{counting}, the  value of the prime counting function $\pi(x)$,  whose fluctuation around its average
value $x/\ln (x)$ is bounded by the Riemann Hypothesis.

Some relevant properties of the prime sequence (and presumably other sequences of numbers and their corresponding quantum states) are encoded  into the density matrix and the quantum correlations among partitions of the Prime state, and are therefore amenable
to investigation using a quantum computer. This approach does not require  the  full  access to the density matrix of the state  
in order to extract  information about the distribution of primes, for that  would cost a number of operations that scales exponentially with the number of qubits. Instead, once the Prime state is constructed,
several number-theoretical functions become direct observables that can be simply measured in the computational basis.
An example is provided by the Chebyshev bias $\Delta(x)$  \cite{Chebyshev}, which is the difference in the number of primes below some $x$
appearing in the arithmetic progressions $4 k +3$ and $4 k +1$, with $k$ integer. 
This bias can be obtained 
by measuring just a single qubit of a Prime state with $\log_2(x)$ qubits \cite{prime}. 

It is specially interesting to note that the Prime state bears an amount of entanglement that is almost maximal across many, if not all, bi-partitions (as characterized by \eg the von Neumann entropy \cite{entangl}). As a consequence, the Prime state is genuinely quantum, that is, it encodes correlations that cannot be described in classical terms. This fact is to be expected. 
Very large entanglement is tantamount to large quantum correlations related to non-locality, and it brings a quantification of surprise among partitions in the system. Digits in a part of the register are almost maximally surprised to relate to those in the rest of the register. If this were not the case, it would be easier to predict the appearance of primes.

Therefore, the main interest in studying the entanglement traits of the Prime state,  and other number-theoretical quantum states,  is that it is not at all unreasonable to think that its quantum correlations may be related to deep facts in Number Theory. For instance, the Hardy-Littlewood constants \cite{H-L} that characterize pairwise correlations among primes appear in the asymptotic expression of the reduced density matrix of the Prime state, for natural bi-partitions \cite{natural}. This fact and others that we shall present in this work can be interpreted as convincing evidence supporting the aforementioned statement. Quantum entanglement may indeed help unravel deep truths in Number Theory.

Let us mention a different line of research that brings together Arithmetics with Quantum Mechanics. The so called {\em coprime chain} \cite{M17} is a one dimensional strongly correlated system where the local degrees of freedom are labelled by integers, and have a nearest-neighbour interaction when they share a common divisor. Another many-body Hamiltonian that has
been constructed recently uses interacting  hard-core bosons on a ladder system \cite{M20}. Here the Prime state emerges as the ground state in a certain limit of the coupling constants. However, in our study we work directly with number-theoretical states and not with Hamiltonians as in Refs. \cite{M17,M20}. Another study that considers quantum states built upon sequences of integers can be found in \cite{sequences}.

In the present work, we report several advances in unveiling properties of the Prime state that relate to Arithmetics. 
First, we introduce the Quantum Fourier Transform (QFT) of the Prime state, which can be efficiently computed and allows to obtain Chebyshev-like biases in the prime number distribution from simple measurements in the computational basis. Then, we study the entanglement present in the Prime state and several of its quantum relatives, up to $n=30$ qubits. We find that the scaling of the entropy for bi-partitions of size $\frac{n}{2}$ is almost identical, and almost maximal, for Prime states written in different qudit basis and for primes in arithmetic progressions. This actual scaling of the entropy is linear in $\frac{n}{2}$ with slope $\sim 0.88\dots$. We conjecture an analytical relation of this slope with the Shannon entropy of the density of square-free integers. A related conjecture is posed regarding the intercept of the entropy.
We provide an analytical approximation to the eigenvalues of the reduced density matrix of the Prime state, for natural bi-partitions. We also find a numerical approximation to the entropy of Prime states with any number of qubits and natural bi-partitions of any size.
Finally, we compute the von Neumann entropy of novel number-theoretical quantum states, defined from the sequences of square-free integers, odd composite numbers and starry primes, that we shall define below. For this study, we have developed an open-source library that diagonalizes matrices using floats of arbitrary precision \cite{github}.

The rest of the paper is organized as follows. Section 2 briefly reviews the Prime state; section 3 introduces its Quantum Fourier Transform; section 4 describes the entanglement properties of the Prime state and other novel number-theoretical quantum states; lastly, section 5 summarizes the conclusions.

\section{Review of the Prime state}

\noindent The Prime state $\P$ of $n$ qubits is defined as the uniform superposition of all the computational-basis states corresponding to prime numbers written in binary format, less than  $2^n$ (we assume that $n \geq 2$),  \ie
\beq \P\equiv \frac{1}{\sqrt{\pi(2^n)}} \sum_{p:\, prime}^{2^n} |p\ra\,, \eeq
\noindent where $\pi(x)$ is the prime-number counting function, which  gives the number of primes smaller than or equal to $x$,   $p=p_{n-1}2^{n-1}+\cdots+p_12^1+p_02^0$, and $|p\ra\equiv|p_{n-1}\ra\otimes\cdots \otimes |p_1\ra\otimes |p_0\ra$. This state was introduced in Ref. \cite{prime}, where it was shown that its construction on a quantum computer is efficient. There are two ways of proceeding. The first simpler method is probabilistic; a second one makes use of Grover's algorithm. Let us review in more detail these two ways of constructing the Prime state. 

The fundamental element to create a Prime state in either a probabilistic or deterministic way is to design a unitary operation $U_{prime}$ that discriminates primes from composites. This unitary acts as follows,
\beq U_{prime}\,|k\ra \otimes|a\ra \equiv \left\{ \begin{array}{ll}
	|k\ra\otimes X\,|a\ra &\;\, \textrm{if $k$ is prime} \\ \\
	|k\ra\otimes |a\ra & \;\,\textrm{if $k$ is not prime} \end{array}\right.\,, \eeq
\noindent 
where $|k\ra$ is a $n$-qubit computational-basis state, $|a\ra$ is a single ancilla qubit, and the action of the $X$ gate is given by the Pauli matrix $\sigma_x$. Note that if the state of the ancilla is $|a\ra=\frac{1}{\sqrt{2}}(|0\ra-|1\ra)$, as done in Grover's subroutine, the action of $U_{prime}$ on a superposition is to introduce relative minus sign for prime numbers. The relevant observation here is that the above unitary amounts to code on a quantum computer a primality test, which is a language that belongs to  the complexity class $P$. As a consequence, the operation $U_{prime}$ only involves a polynomial number of quantum gates, which will depend on the specific primality test chosen. For the sake of illustration, a detailed explicit form of a quantum primality oracle with $O(n^6)$ gates, based on the classical Miller-Rabin test \cite{Miller-Rabin}, was produced in Ref. \cite{prime}. Other primality algorithms, such as the AKS primality test \cite{AKS}, could be used as well. 

We can  then consider a first probabilistic way to create the Prime state.
We need to apply the unitary $U_{prime}$ on the uniform superposition of all computational-basis vectors, $|\phi\ra=\frac{1}{\sqrt{2^n}}\,\sum_{i=0}^{2^n-1}|e_i\ra$, plus an ancilla in the $|0\ra$ state; where $|\phi\ra$ is created by applying $n$ Hadamard gates, $H^{\otimes n}$, onto the initial $|0\dots0\ra$ state. This yields:
\beq \label{probabilistic} \sqrt{\frac{2^n-\pi(2^n)}{2^n}} \sum_{\overline{p}:\,not\, prime} |\overline{p}\ra|0\ra +\sqrt{\frac{\pi(2^n)}{2^n}} \sum_{p:\,prime}|p\ra|1\ra\,.  \eeq
By measuring the ancillary qubit in the above state, and post-selecting whenever the result of the measurement is $|1\ra$, one obtains the Prime state. This will occur with probability $\sim\frac{1}{n\ln\,2}$, according to the Prime Number Theorem (see below). Because this probability decreases only polynomially with $n$, this is an efficient method to create the state. 

A more refined, deterministic way to create the Prime state consists in using a Grover's algorithm that uses a primality test as oracle. 
The efficiency of the algorithm hinges on two facts.  The first was mentioned previously, that is, the oracle based on $U_{prime}$ is efficient. The second fact that guarantees the efficiency in the construction of the Prime state is the relatively high abundance of primes, $\pi(x)$, which is asymptotically given by the 
logarithmic integral function, $Li(x)\equiv\int_2^x\frac{dt}{\ln\,t}$, \ie
\beq \label{PNT} \pi(x)\sim Li(x)\;\xrightarrow{x\rightarrow \infty} \;\frac{x}{\ln\,x}\,. \eeq
This result is known as the Prime Number Theorem (PNT) \cite{NumberTheory}, and it implies that the needed number of calls to Grover's oracle is only $O(\sqrt{n})$. Indeed, this estimate is given by $\frac{\pi}{4}\,\sqrt{\frac{N}{M}}$ \cite{Chuang}, where $N=2^n$ is the dimension of the Hilbert space of $n$ qubits and $M$ is the number of solutions to the oracle, \ie $M=\pi(2^n)$. Therefore, the overall computational complexity of generating a Prime state of $n$ qubits on a quantum computer with this method, using \eg Miller-Rabin primality test, is $O(n^{6.5})$ \cite{prime}.

As mentioned previously, once an oracle for the Prime state has been created, it can be used to numerically verify (or falsify) 
the Riemann hypothesis, which constitutes one of Clay Mathematics Institute's Millennium Problems \cite{Millennium}. 
The Riemann hypothesis states, for the distribution of prime numbers, that the deviations of $\pi(x)$ from $Li(x)$ 
are bounded asymptotically with $x$ as  \cite{2657}
\beq \label{riemann} |\pi(x)-Li(x)|\leq \frac{1}{8\pi} \sqrt{x}\ln{x} \qquad\textrm{for $x>2657$}\,.\eeq
The quantum algorithm proposed in Ref. \cite{prime} allows to compute $\pi(x)$ beyond the limits achievable by means of classical computation --once a fault-tolerant universal quantum computer is built-- by using a quantum counting algorithm that provides an estimate of the number of terms in superposition in the Prime state. So far, $\pi(x)$ have been computed up to $10^{27}$ \cite{pi(x)}, which implies that a Prime state with a minimum of $\sim90$ logical qubits would be needed to surpass this computation (since $2^{90}\simeq 1.238\times 10^{27}$).

To do so, it was suggested to apply a quantum counting algorithm \cite{counting} that delivers the number of solutions $M$ to an oracle search problem; in this case, the number of primes that are marked by Grover's oracle, $G$. This algorithm uses Quantum Phase Estimation (QPE) \cite{QPE} to obtain the eigenvalues of $G$, which in turn reveal the number of solutions to the query problem. In order to obtain an estimate of $\pi(x)$ that is meaningful, the precision in the estimation should be smaller than the fluctuations allowed by the Riemann Hypothesis. To achieve such precision, $O(\sqrt{2^n})$ calls to the oracle are needed. This is quadratically better than the performance of any classical counterpart in an identical oracular setting of the counting problem. Recently, there have been new proposals for quantum counting the solutions to an oracle $G$ that do not rely on QPE \cite{aaronson}. In practice, any quantum counting algorithm may be applied.

For the sake of completeness, let us recall that the best classical algorithms that compute $\pi(x)$ unconditionally, \ie the validity of the estimation not depending on the truthfulness of the Riemann Hypothesis or any other unproven conjecture, do not however rely on enumerating all primes. 
The main algorithm for computing $\pi(x)$ is a combinatorial method due to Meissel \cite{Meissel} and Lehmer \cite{Lehmer}, which has subsequently been improved by Lagarias, Miller and Odlyzko \cite{LMOdlyzko}, Deléglise and Rivat \cite{DR}, and Gourdon \cite{Gourdon}. The latter version of this algorithm, which has time complexity $O(x^{2/3}\ln^{-2} x)$, was used to compute $\pi(10^{27})$.
Another two analytic algorithms for computing $\pi(x)$, put forward by Lagarias and Odlyzko \cite{LagariasOdlyzko}, have time complexity $O(x^{1/2+\epsilon})$ and $O(x^{3/5+\epsilon})$ respectively, with $\epsilon>0$.

\section{Quantum Fourier Transform of the Prime state}

\noindent The Quantum Fourier Transform (QFT) is a unitary operation that plays a key role in several important algorithms, such as Quantum Phase Estimation \cite{QPE} or Shor's algorithm \cite{Shor}. Its action on a quantum state $\ket=\sum_j x_j\,|e_j\ra$ of $n$ qubits, expressed in the computational basis, $\{|e_j\ra\}$, is given by
\beq \label{QFTdef} QFT \,|\psi\ra = \sum_{k=0}^{2^n-1}y_k\,|e_k\ra\;,\quad y_k\equiv\frac{1}{\sqrt{2^n}} \sum_{j=0}^{2^n-1} x_j\,e^{2\pi i\,jk/2^n}\,,  \eeq
where the $\{y_k\}$ are the discrete Fourier transform of the original amplitudes $\{x_j\}$, and the $i$ in the exponent is the imaginary unit. The QFT is an efficient subroutine, requiring $O(n^2)$ gates on a quantum computer \cite{Chuang}. 

\begin{figure}[t!]
	\centering
	\includegraphics[scale=0.6]{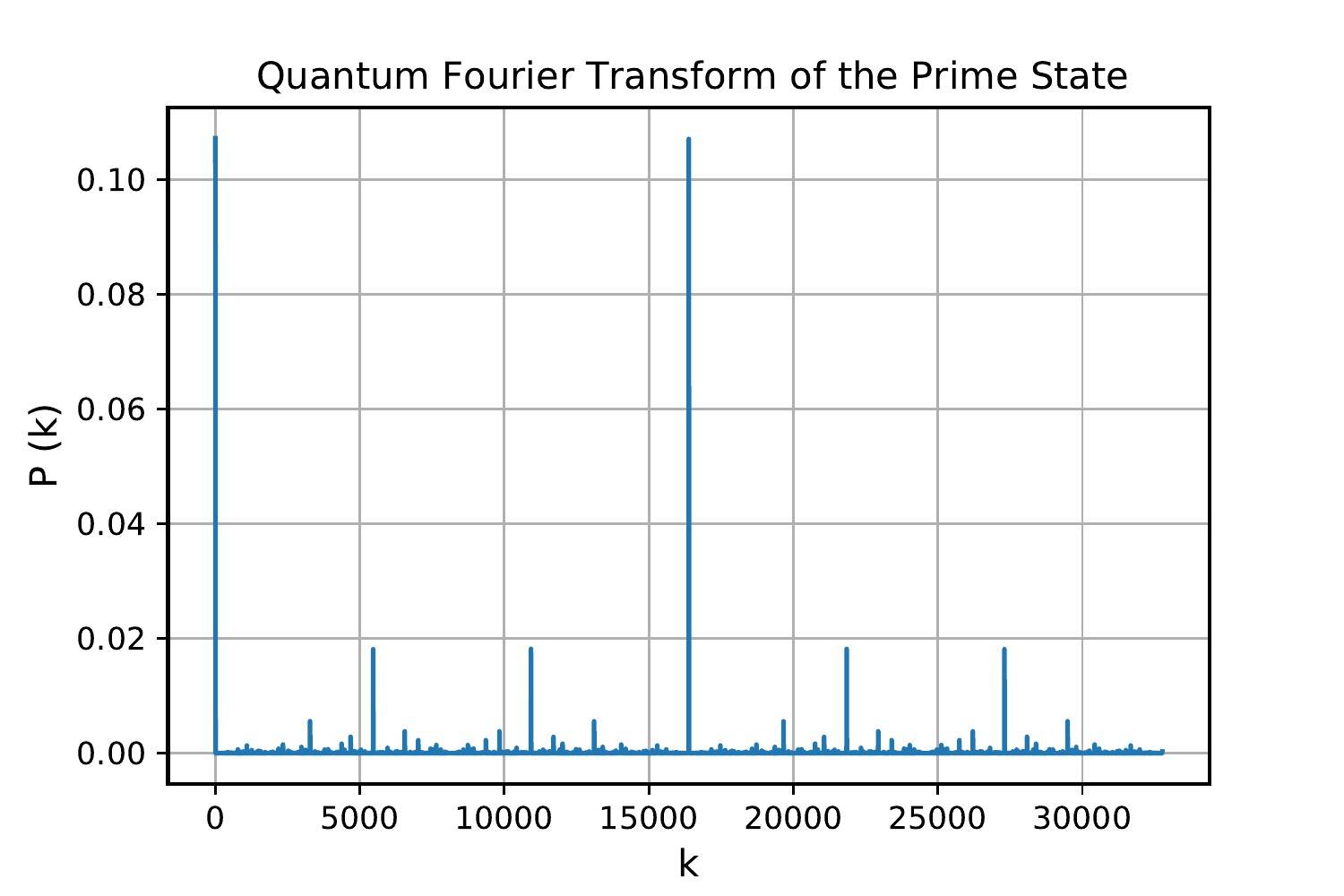}
	
	\caption{\small{Numerical simulation of the Quantum Fourier Transform of a Prime state with $n=15$ qubits. The probability peaks appear symmetrically distributed with respect to the central value $k=2^n/2$. Note as well that there is no peak for the value $k=2^n$, because the state  $|2^n\ra$ is not included in the Hilbert space of $n$ qubits.}}
	\label{fig:qft}
\end{figure}

In the present work, we calculate the QFT of the Prime state, $|\hat{\mathbb{P}}_n\ra\equiv QFT \,\P$, that is,
\beq \label{QFT} |\hat{\mathbb{P}}_n\ra=\frac{1}{\sqrt{2^n\pi(2^n)}} \sum_{k=0}^{2^n-1} \left(\sum_{p:\,prime} e^{2\pi i\,pk/2^n}\right)|k\ra\,.\eeq
Notice that the amplitudes in Eq. \eqref{QFT} are symmetric with respect to the central value $k=N/2$, where $N=2^n$. This means that the probability of measuring the state $|k\ra$ and $|2^n-k\ra$ is the same. Indeed:
\ba P(2^n-k) & =  & \frac{1}{2^n\,\pi(2^n)}\left| \sum_{p:\,prime}e^{2\pi i\,p}\,e^{-2\pi i\,pk/2^n} \right|^2  \\ 
& = &  \frac{1}{2^n\,\pi(2^n)}\left| \sum_{p:\,prime}e^{-2\pi i\,pk/2^n} \right|^2 = \,P(k)\,.  \nonumber 
\ea
This is a general property of the QFT of a state with real amplitudes, 
that is verified by the Prime state, see Fig. \ref{fig:qft}.

\bigskip

The main result of the computation of $|\hat{\mathbb{P}}_n\ra$ is that the QFT applied to the Prime state provides a method for estimating  Chebyshev-like biases in the distribution of prime numbers. These biases reflect the unbalance in the number of primes, below a certain value, appearing in different arithmetic progressions.

Let us therefore consider arithmetic progressions of the form $\alpha k +\beta$, for $k=0,1,2,\dots$, with $\alpha,\beta$ coprimes (\ie $\gcd(\alpha,\beta)=1$). Dirichlet proved that there exists an infinite number of primes in any of these sequences \cite{NumberTheory}. The number of primes in the sequence $\alpha k +\beta$ (with $\alpha,\beta$ coprimes), below a certain value $x$, is denoted by the modular prime counting function $\pi_{\alpha,\beta}(x)$, in close analogy to the prime counting function, $\pi(x)$. The asymptotic behaviour of these modular counting functions is determined by
\beq \label{pi,ab} \pi_{\alpha,\beta}(x)\sim \frac{1}{\phi(\alpha)}\,Li(x)\;\xrightarrow{x\rightarrow\infty}\;\frac{1}{\phi(\alpha)}\frac{x}{\ln\,x} \,,\eeq
where $\phi(\alpha)$ is the Euler's totient function, which gives the number of coprimes to $\alpha$ smaller than $\alpha$. 
For example, if $p$ is a prime, one has $\phi(p) =p-1$, because  all the integers below $p$ do not divide it. Notice as well that $\phi(\alpha)$ is the number of arithmetic progressions of the form $\alpha k +\beta$ that can contain primes, which are those in which $\beta$ is coprime to $\alpha$. Thus, Eq. \eqref{pi,ab} means that the number of primes is, on average, asymptotically equi-distributed among the existing progressions $\alpha k+\beta$ ($\alpha,\beta$ coprimes), for fixed $\alpha$. 
Nevertheless, when considering a finite natural  sequence of primes,
there exists  biases in the distribution of the primes among different  progressions that are quantified by the functions
\beq \Delta_{\alpha;\,\beta_1,\beta_2}(x)\equiv \pi_{\alpha,\beta_1}(x)-\pi_{\alpha,\beta_2}(x)\,.\eeq
These biases can be numerically estimated, to a desired precision $\epsilon$, from the probability peaks appearing in $|\hat{\mathbb{P}}_n\ra$. In particular, relevant peaks appear at values $k=N/3,N/4,N/6$, and their mirror images 
about the central peak (see Appendix A for a derivation of these results). The expressions for these peaks are,
\beq \label{peaks} \begin{split} P(N/3)&= \frac{\Delta_{3;2,1}(N)^2+\pi_{3,1}(N)\pi_{3,2}(N)-\pi(N)+2}{N\pi(N)}\,,  \\\\ P(N/4)&=\frac{1+\Delta(N)^2}{N\pi(N)} \,,\\\\ P(N/6)&= \frac{\Delta_{6;5,1}(N)^2+\pi_{6,1}(N)\pi_{6,5}(N)-3\,\pi_{6,1}(N)+3}{N\pi(N)}\,, \end{split}\eeq
where $\Delta(N)\equiv \pi_{4,3}(N)-\pi_{4,1}(N)$ is the Chebyshev bias. There exists  a direct relation between the functions $\pi_{3,1}(x)$, $\pi_{3,2}(x)$ and $\pi_{6,1}(N)$, $\pi_{6,5}(N)$. In fact, $\pi_{3,1}(x)=\pi_{6,1}(x)$ and $\pi_{3,2}(x)=\pi_{6,5}(x)+1$. This is so because every prime $p$ counted by $\pi_{3,1}(x)$ is of the form $p=3k+1$, but since $p$ is odd, $k$ must be even ($k=2k'$), and $p=6k'+1$, so $\pi_{3,1}(x)=\pi_{6,1}(x)$. On the other hand, every prime $p$ counted by $\pi_{3,2}(x)$ is of the form $p=3k+2$, with $k$ odd ($k=2k'+1$), so $p=6k'+5$;  in this case, 2 is the only prime counted by $\pi_{3,2}(x)$ and not by $\pi_{6,5}$, and thus $\pi_{3,2}(x)=\pi_{6,5}(x)+1$. These relations,
\beq\label{relations} \pi_{3,1}(x)=\pi_{6,1}(x)\quad,\quad\pi_{3,2}(x)=\pi_{6,5}(x)+1 \,, \eeq
allow to obtain $\pi_{3,1}(x)$, $\pi_{3,2}(x)$, $\pi_{6,1}(N)$, $\pi_{6,5}(N)$, $\Delta_{3;\,2,1}(N)$ and $\Delta_{6;\,5,1}(N)$, by substituting Eq. \eqref{relations} into the probability peaks $P(N/3)$ and $P(N/6)$ in Eq. \eqref{peaks}, and solving the resulting system of two equations and two unknowns. Of course, $\pi(N)$ must be known in advance, but this is already provided by the Prime state combined with some quantum counting algorithm. Notice as well that $N/3$ and $N/6$ do not correspond to integer values for qubit systems, where $N=2^n$. This means that the associated peaks cannot be directly measured in the computational basis. Instead, one needs to resort to ancillary qubits to represent binary fractions and hence rational numbers; the QFT acts upon the two registers, the one representing the integer part and the one representing the fraction (see Appendix A for a numerical check of the error 	incurred using nine ancillary qubits).

It is important to consider how efficient it is to measure the peaks in Eq. \eqref{peaks}, to a desired precision, when $n$ grows large.
Note that $P(N/3)$ and $P(N/6)$ are dominated by the cross product of the modular prime counting functions, divided by $N\pi(N)$. According to Eq. \eqref{PNT} and Eq. \eqref{pi,ab}, these probabilities will then decrease as $\sim\frac{1}{\phi(\alpha)^2}\frac{1}{n\ln\,2}$, where $\alpha=3,6$ respectively. This decrease is only polynomial in $n$, and therefore, the method is efficient. This is so because the number of measurements required to achieve a target precision in the estimation of outcome probabilities, $\{p_i\}$, is dictated by the sampling process of a multinomial distribution. In this context, the statistical additive errors \cite{errors}, $\{\epsilon_i\}$, arise from finite sampling, and depend on the number of measurements performed, $\#M$, as $\epsilon_i=\sqrt{\frac{p_i(1-p_i)}{\#M}}$. Hence, in order to achieve a small multiplicative error $\varepsilon_i$, say $\varepsilon_i\approx0$ and $\epsilon_i<<p_i$, then the number of measurements, or runs of the quantum circuit, must be $\#M>>\frac{1-p_i}{p_i}$. Therefore, it is clear that estimating probabilities with a small multiplicative error remains efficient as long as these probabilities show at most a polynomial decrease with increasing $n$.
 
Moreover, direct application of the QFT on the Prime state, followed by the procedure of Amplitude Estimation \cite{AE}, allows to obtain a dependency of the additive error on the number of runs of the quantum circuit that is $O(\frac{1}{\# M})$, \ie a quadratic improvement compared to direct sampling from the multinomial distribution. This is the same technique employed for instance in the quantum Monte Carlo algorithm \cite{MonteCarlo}. In this case, all it takes to prepare $|\hat{\mathbb{P}}_n\ra$ for Amplitude Estimation is to apply a single $n+1$-qubit controlled-$X$ gate on an ancilla qubit, that singles out the desired peak, \eg $|N/3\ra$,
\beq \label{AE} \sum_{\substack{k=0\\k\neq N/3}}^{N-1}y_k\,|e_k\ra |0\ra + \sqrt{P(N/3)}\,|N/3\ra|1\ra\,. \eeq
States like the one in Eq. \eqref{AE} are ready for Amplitude Estimation. 
Classically, the best algorithm for computing modular prime counting functions, $\pi_{\alpha,\beta}(x)$, is a variation of the Meisser-Lehmer method for computing $\pi(x)$, and has time complexity $O(x^{2/3}\ln^{-2}x)$ \cite{ModularCounting}. To the best of our knowledge, modular prime counting functions have been computed for values only up to $x=10^9$ \cite{wolfram}. Hence,  a quantum computer could be helpful in this sort of computations.

In contrast to $P(N/3)$ and $P(N/6)$, the peak $P(N/4)$, which delivers the Chebyshev bias, $\Delta(N)$, is unlikely to be of practical use because the denominator $N\pi(N)$ is expected to cause an exponential decrease in this probability, as the number of qubits $n$ increases. As already explained, this would demand an exponentially large number of measurements. Yet, it is interesting to appreciate how the amplitudes of the Prime state interfere under the QFT to give some remarkable meaning, in terms of number-theoretical functions, to certain values of $k$.
Note as well that if one is willing to directly sample from $|\hat{\mathbb{P}}_n\ra$, all the information contained in the probability peaks is extracted {\sl simultaneously}, rather than sequentially, by accumulation of statistical knowledge on measurement outcomes. This means that the values of {\sl all} peaks are estimated altogether with a fixed number of samples,  to a desired precision $\epsilon$.

It also happens that the QFT of an equally-weighted superposition provides a means to count the number of terms in the computational basis. Indeed, the probability of finding the $|0\dots0\ra$ state on a measurement after the application of the QFT equals $M/N$, where again $N=2^n$ is the dimension of the Hilbert space of $n$ qubits and $M$ is the number of terms in the superposition (in the computational basis). A derivation of this result can be found in Appendix B. For $|\hat{\mathbb{P}}_n\ra$, this probability reads
\beq P(0)= \frac{\pi(N)}{N}\,, \eeq
which decreases as $\sim\frac{1}{n\ln\,2}$ in the limit $n \gg 1$,  according to the PNT, Eq. \eqref{PNT}.  The prime counting function, $\pi(N)$, appears as well in the central peak, where $k=N/2$, as shown in Appendix A. The expression for this peak is
\beq P(N/2)=\frac{\pi(N)^2-4\pi(N)+4}{N\pi(N)} \,,\eeq
which also decreases linearly in $n^{-1}$.

Because the precision required to meaningfully test Riemann's hypothesis, Eq. \eqref{riemann}, implies that the multiplicative error must become smaller and smaller for increasing $x$ (as $\frac{\sqrt{x}\,\ln\,x}{\pi(x)}$ decreases), direct sampling from $|\hat{P}_n\ra$ is not competitive with previously mentioned quantum counting algorithms, requiring $O(2^{n\,} n^{-3})$ repetitions. 
On the other hand, it is possible to prepare a state like that in Eq. \eqref{AE}, but marking the peak $P(0)$ instead, and then apply Amplitude Estimation. However, note that if one has an oracle, it is possible to generate, with a single query to that oracle, the state in Eq. \eqref{probabilistic}. Such state is also ready for AE, and can be used to estimate $\pi(N)=M/N$, but it is easier to prepare than the one in Eq. \eqref{AE}. So there is no point in using $|\hat{\mathbb{P}}_n\ra$ for estimating $\pi(2^n)$.

The QFT may nonetheless be useful to estimate the number of terms in an equally-weighted superposition, for states for which one does not know of any oracle; for instance, a ground state of a Hamiltonian obtained from a variational algorithm such as the Variational Quantum Eigensolver \cite{vqe} or the Quantum Approximate Optimization Algorithm \cite{qaoa}.

\section{Entanglement of number-theoretical quantum states}

\noindent The entanglement properties of number-theoretical quantum states can be studied by taking different bi-partitions of the states and quantifying their entanglement, with figures of merit such as the von Neumann entropy, the purity or the entanglement spectrum, among many possibilities. Unfortunately, a practical way to quantify {\sl genuine} multipartite entanglement, not related to partitions of the system, is not available for states with a large number of qubits \cite{genuine}. 

\subsection{The Prime state}

The entanglement present in number-theoretical quantum states is the result of correlations among the digits of the numbers appearing in the superposition. In turn, these observables may reveal information about pairwise correlations between the numbers in the sequence. For instance, the reduced density matrix of the Prime state upon taking a natural bi-partition $A$--$B$ with the first (\ie least significant) $m$ qubits \cite{natural}, $\rho_A$, is asymptotically given by $\overline{\rho}_A$ \cite{entangl}, whose expression is
\beq \label{densmatrix} \overline{\rho}_A=\frac{1}{d}\left({\bf \mathbbm{1}}+\ell_N{\bf C}_m\right)\,,\eeq 
where $d=2^{m-1}$, $\ell_N\equiv\frac{Li_2(N)}{Li(N)}\xrightarrow{N\rightarrow \infty}\frac{1}{n\ln\,2}$, and ${\bf C}_m$ is a $d\times d$ symmetric Toeplitz matrix defined by
\beq \label{HLconstants} ({\bf C}_m)_{i,j}\equiv (1-\delta_{i,j})\,C(2|i-j|), \quad  i, j =1, \dots, d \, , 
\eeq
 where $C(h)$ are the  Hardy-Littlewood constants, defined by
\beq 
\label{HL}
C(h)=\left\{ \begin{array}{lc}
	2\, C_2\,\prod_{p>2;\,p|h} \frac{p-1}{p-2}&  \textrm{ if $h$ is even} \\\\ 0 &  \textrm{otherwise}
\end{array}\right.\,,\eeq 
with $C_2=\prod_{p>2}\left(1-\frac{1}{(p-1^2)}\right)$ the twin prime constant and $\{p\}$ the prime numbers.  
The constants $C(h)$  characterize the pairwise correlations among prime numbers, 
and it is precisely due to these correlations that the reduced density matrix $\overline{\rho}_A$ 
is not maximally mixed. To be precise, the first Hardy-Littlewood conjecture \cite{H-L} assures that the number of prime pairs $(p,p+k)$ up to a certain number $x$, \ie $\pi_k(x)$, is given by
\beq \pi_k(x)\sim C(k)\, Li_2(x)\;\xrightarrow{\,x\rightarrow\infty\,}\; C(k)\,\frac{x}{(\ln\,x)^2}\,,
\label{HLC}
\eeq
where $Li_2(x)=\int_2^x\frac{dt}{(\ln\,t)^2}$.

An analytical approximation to the positive eigenvalues of the matrix ${\bf C}_m$ was conjectured  in Ref. \cite{entangl}. 
That approximation was suitable for the highest eigenvalues, but failed to describe the lowest ones. 
Moreover, the degeneracies of the former failed beyond some point \cite{Botero}.
A recent article gives   a heuristic derivation of the Hardy-Littlewood conjecture \eqref{HLC}
starting from  the pair correlation formula for the Riemann zeros, including lower order terms \cite{Keating}. 
The reverse statement was known since long and played   an important role in the connection between
Number Theory and Quantum Chaos \cite{BK96,BK99}. The derivation of Ref. \cite{Keating} 
employs the following expression  for the Hardy-Littlewood constants,
\barray
C(h)    
& =  &     
\sum_{k=1}^\infty  \left( \frac{ \mu(k)}{ \phi(k)} \right)^2  c_k(h) 
 \label{K54}
\earray 
where $\phi(x)$ is the Euler's totient function, $\mu(x)$ is the M\"obius function (see below), and
\barray
c_k(h)    
& =  &    \sum_{\substack{l=1\\ {\rm gcd}(l,k)=1}}^k   
  e^ {2 \pi i \,h l/k}  
\label{K55} 
\earray 
are the  Ramanujan's sums.   Equation \eqref{K54}  is  the Ramanujan-Fourier
series of the constants $C(h)$  and has also  appeared  in the context of random processes, 
more concretely,  in the relation between the power spectrum and the correlation function 
using the  Wiener-Khintchine formula  
\cite{GP99,GP06}. 


\begin{figure}[t!]
	\centering
	\includegraphics[scale=0.6]{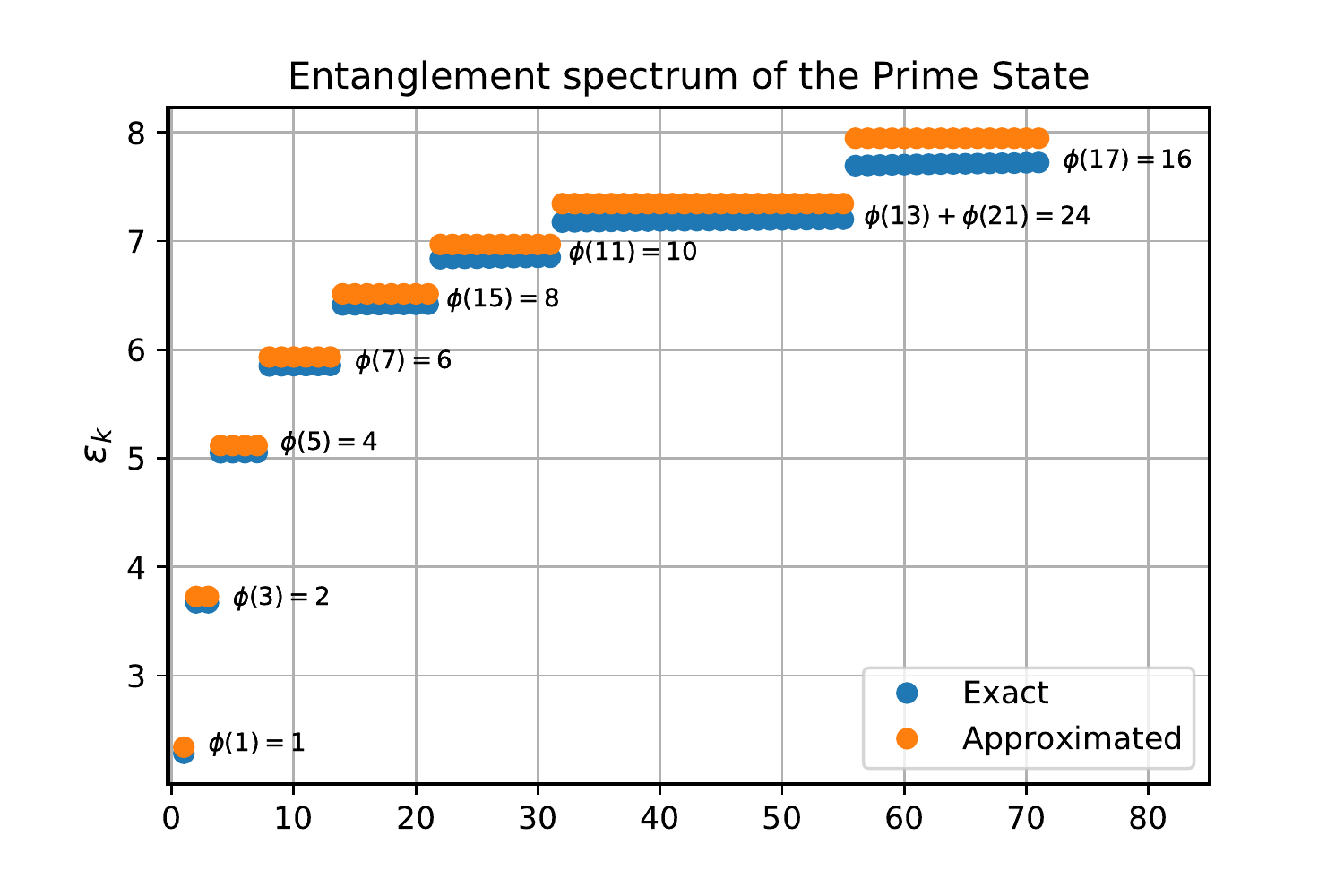}
	
	\caption{\small{Comparison of the exact lowest 72 entanglement energies of the entanglement spectrum of the Prime state to those computed using Eq. \eqref{Eigvals}, for $n=30$ qubits. We expect this approximation  to improve for $n \gtrsim 64$,
	based on the very good approximation obtained for the slope of the entanglement entropy, Eq. \eqref{c_pi}, shown in Fig. \ref{diffS}-right in Appendix C. }} 
	\label{fig:eigenvalues}
\end{figure}

Based on  Eq. \eqref{K54}, a more precise description of the eigenvalues of ${\bf C}_m$ can be obtained
(for a justification see Appendix  C). 
To every square-free  odd integer $k$, that is,  to every odd integer $k$ such that it does not contain any prime raised to a power larger than 1 in its prime decomposition, we associate an eigenvalue $\gamma_k $ of ${\bf C}_m$ with degeneracy $\phi(k)$. These eigenvalues are given by
\beq
\gamma_k \simeq 2^m \,\mu^2(k) \left( \frac{1}{ \phi^2(k)} - \frac{1}{\phi_m}  \right), \quad k =1, 3, 5,  \dots, k_m \,,
\label{eigenvalues}
\eeq
where the M\"obius function, $\mu(x)$, takes values $\pm1$ for square-free numbers and 0 otherwise. The values of $k_m$ and $\phi_m$ are fixed by the dimension and normalization of the density matrix in Eq. \eqref{densmatrix}, and $k_m$ scales as $2^{m/2}$
(see Appendix C for details). There are cases where the eigenvalue is the same for two different values of $k,k'$, in which case
the degeneracy is higher. An example of such an accidental degeneracy is $\widehat{\gamma}_{13} = \widehat{\gamma}_{21} = \frac{1}{144}$,  
which has a total degeneracy $\phi(13)+ \phi(21) = 24$ (we denote by $\widehat{\gamma}_{k} = \mu^2(k)/\phi^2(k)$). Another example is 
$\widehat{\gamma}_{35} = \widehat{\gamma}_{39} = \frac{1}{576}$, that has a degeneracy
$\phi(35)+ \phi(39) = 48$.

Using Eq. (\ref{eigenvalues}), the eigenvalues of the density matrix (\ref{densmatrix}) are approximated by 
\beq
\lambda_k \simeq 2^{ 1- m} \, \mu^2(k)  \left( 1 + \frac{2^m }{n \ln\, 2}\left(\frac{1}{\phi^2(k)}-\frac{1}{\phi_m}\right)      \right) \,,
\label{Eigvals}
\eeq
with a degeneracy $\phi(k)$. The good agreement between the exact eigenvalues and those computed from Eq. \eqref{Eigvals} for $n=30$ qubits, is shown in Fig. \ref{fig:eigenvalues}, where the lowest 72 entanglement energies of the entanglement spectrum are plotted. 
These correspond to the highest eigenvalues of the density matrix of the Prime state, 
since the entanglement spectrum $\{\varepsilon_k\}$ of a density matrix $\rho$ with eigenvalues $\{\lambda_k\}$ is defined as
\beq \varepsilon_k = -\ln\, \lambda_k\,. \eeq

\begin{figure}[t!]
	\centering
	\includegraphics[scale=0.6]{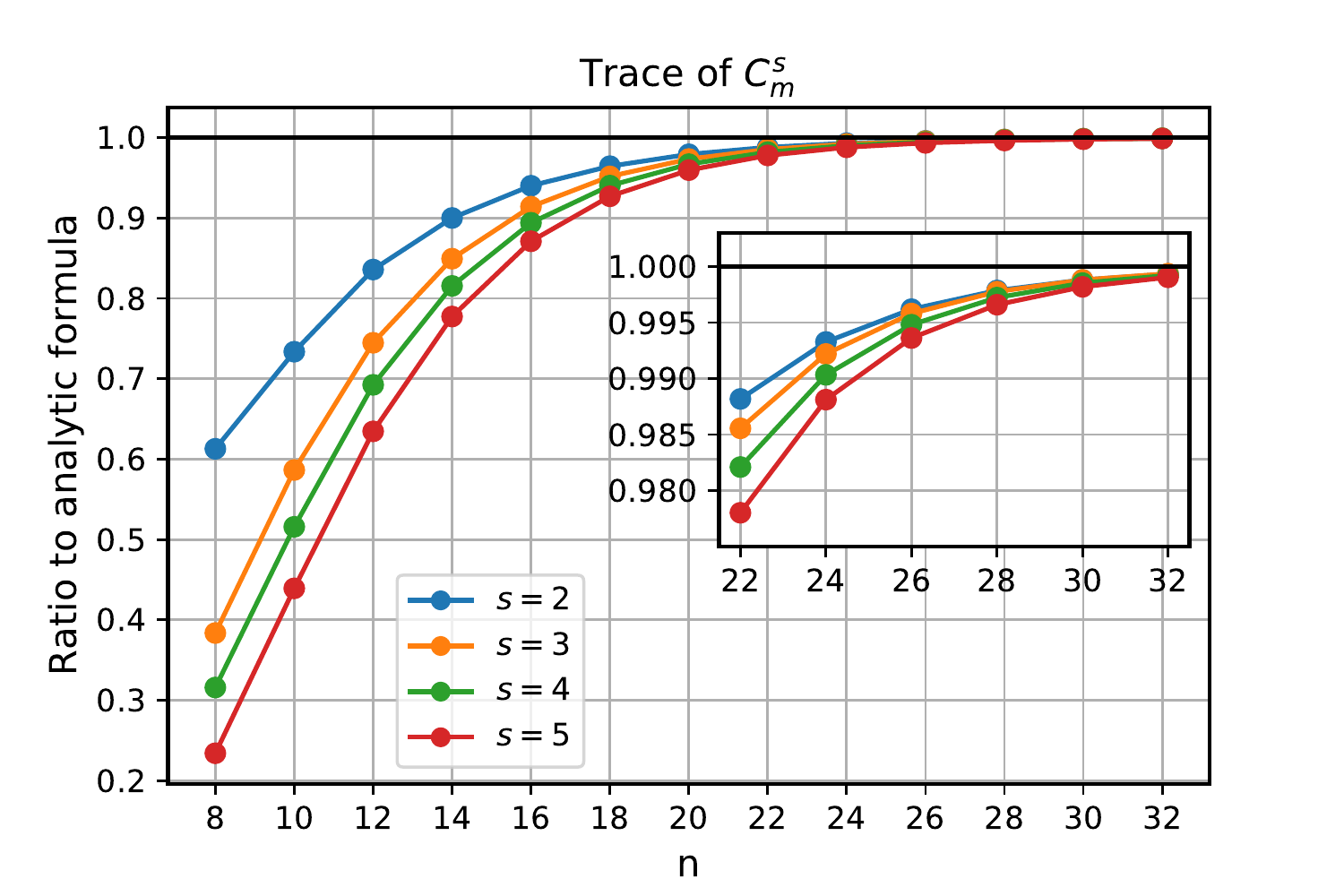}
	
	\includegraphics[scale=0.6]{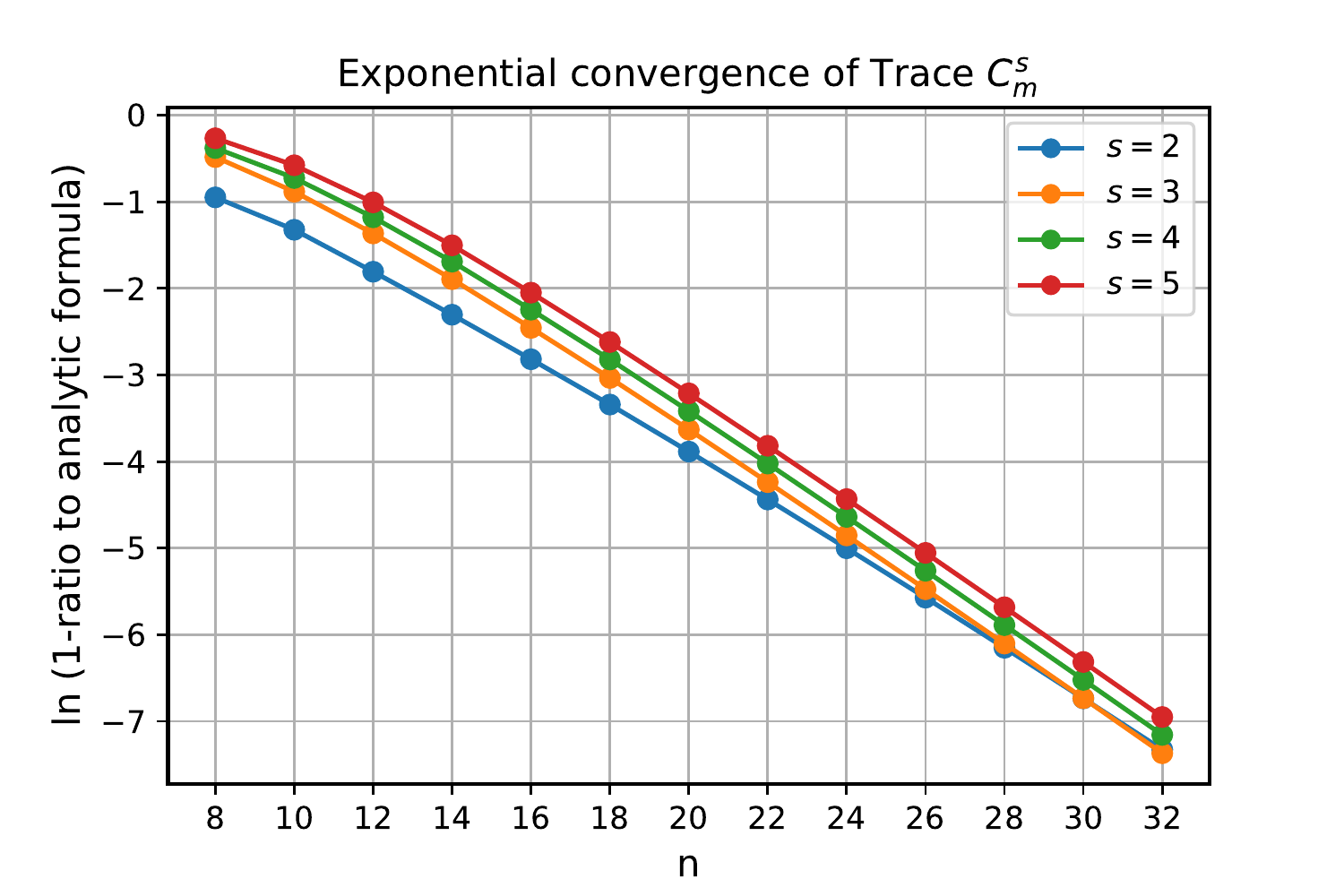}
	
	\caption{\small{top) Ratio of ${\rm Tr}\left({\bf C}_m^s\right)$ to Eq. \eqref{trace}, for $s=2,3,4,5$} and $m=\frac{n}{2}$, up to $n=32$ qubits. bottom) Natural logarithm of ($1-\,$ratio of ${\rm Tr}\left({\bf C}_m^s\right)$ to Eq. \eqref{trace}), up to $n=32$ qubits.}
	\label{fig:traces}
\end{figure}

Using again Eq. \eqref{eigenvalues}, we can estimate the contribution of the positive eigenvalues to the trace of the powers of 
${\bf C}_m$, from which all Rényi entropies can be readily derived \cite{Renyi}. The computation is as follows,
\begin{eqnarray} \label{trace}
{\rm Tr} \; {\bf C}_m^s &  \simeq &   \sum_{k =1 \,({\rm odd}) }^{k_m}   \phi(k) \, \gamma_k^s  \nonumber \\
& \xrightarrow{\,m\gg 1\,} &
2^{ ms}  \sum_{k =1 \,({\rm odd}) }^{\infty}   \frac{ |\mu(k)|}{(\phi(k))^{ 2 s-1}} \nonumber\\ & = & 2^{ms}
\prod_{p > 2} \left( 1 +  \frac{ |\mu(p)|}{(\phi(p))^{ 2 s-1}}  \right) 
\nonumber  \\
& = &  2^{ms}
\prod_{p > 2} \left( 1 +  \frac{ 1}{(p-1)^{ 2 s-1}}  \right) 
 \,,
\end{eqnarray}
where we have used that for any  function $f(k)$ which is multiplicative, it holds that
\beq
\sum_{k =1 \,({\rm odd}) }^{\infty} |\mu(k)|\,   f(k)  = \prod_{p > 2} \left( 1 +  f(p)   \right) \,,
\label{multiplicative}
\eeq
where the product runs  over the odd  prime numbers. We have also used that $\phi(p)=p-1$ for prime numbers, and that $\mu(k)^2=|\mu(k)|$. 
In Eq. \eqref{trace} we dropped  the negative  term proportional to $1/\phi_m$ of  Eq. \eqref{eigenvalues}  because its 
contribution vanishes  in the limit $N,m \gg1$ for $s >1$. 
For $s=2$,  Eq. \eqref{trace} coincides with  the heuristically-derived asymptotic expansion of ${\rm Tr} \; {\bf C}_m^2$, 
and was conjectured to hold for any $s\geq2$ in Ref. \cite{entangl}. A direct connection thus exists between Eq. \eqref{eigenvalues} and the latter conjecture. For $s=1$, the Riemann zeta function $\zeta(1)$ arises and the product diverges because we have dropped the negative eigenvalues of the matrix ${\bf C}_m$. 
Figure \ref{fig:traces} shows the ratio of the exact value of the trace of the matrix ${\bf C}_m^s$ to the asymptotic analytical formula Eq. \eqref{trace}, and the exponential convergence of this ratio towards 1, up to $n=32$ qubits. These results indicate that the contribution of the negative eigenvalues of ${\bf C}_m$ to the trace of the powers of this matrix is asymptotically negligible, for $s\geq2$. In contrast, their contribution to the von Neumann entropy does not appear to be so.

\begin{figure}[t!]
	\centering
	\includegraphics[scale=0.6]{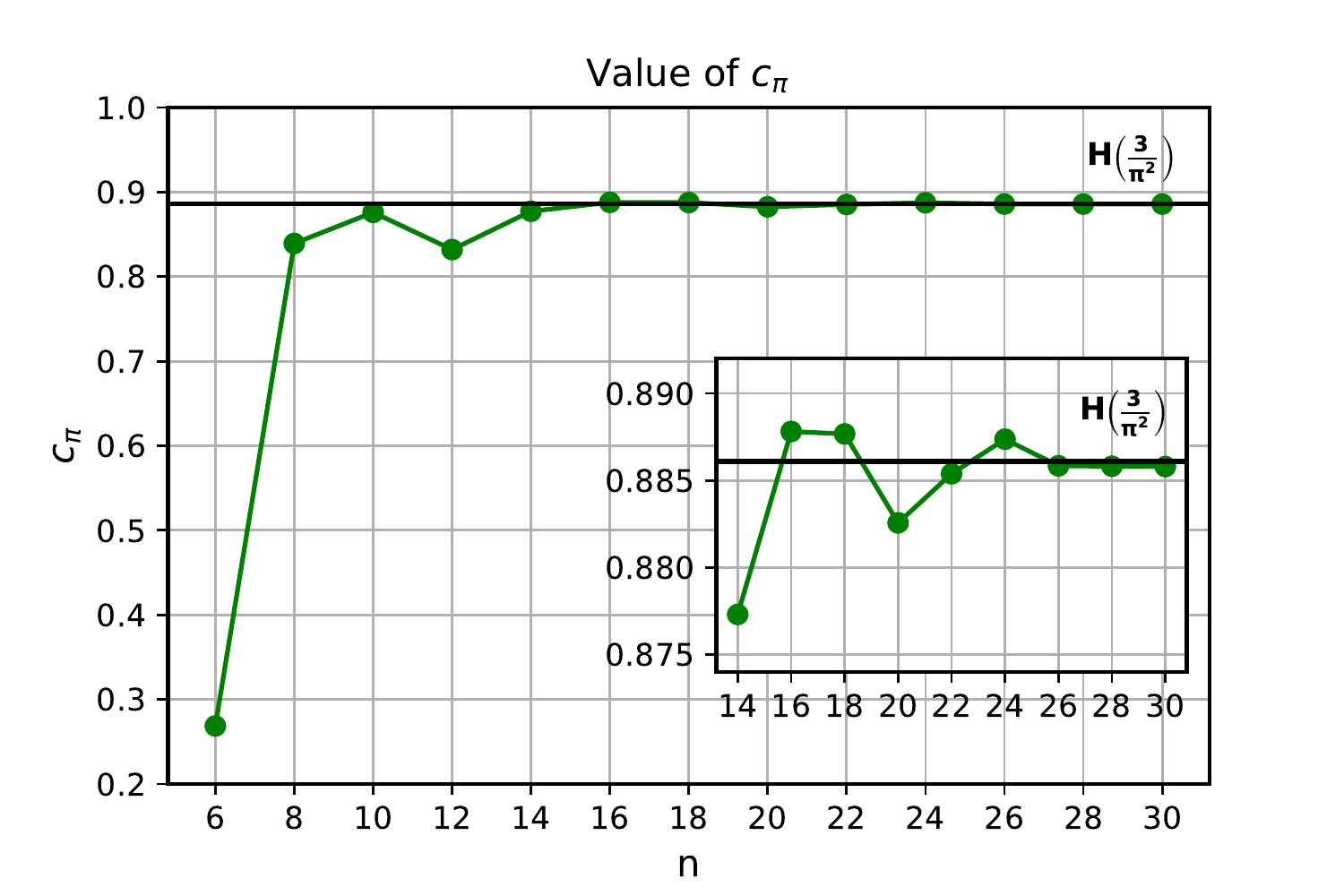}
	
	\includegraphics[scale=0.6]{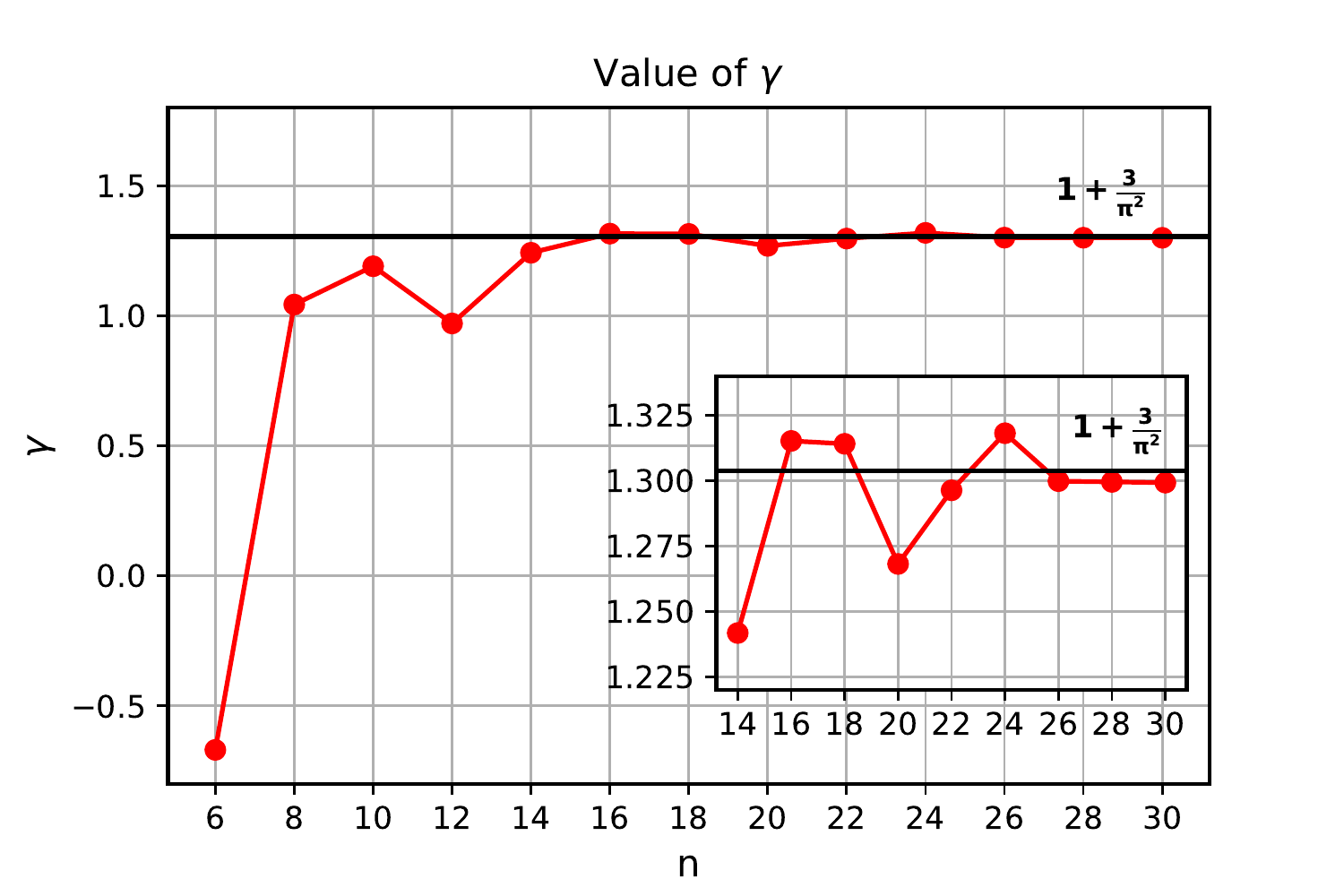}
	
	\caption{\small{top) Value of $c_\pi(n)$ computed according to Eq. \eqref{c_pi}, compared to the conjectured value $H\left(\frac{3}{\pi^2}\right)$, up to $n=30$ qubits; bottom) Value of $\gamma(n)$ computed according to Eq. \eqref{c'}, up to $n=30$ qubits}, compared to the conjectured value $1+\frac{3}{\pi^2}$.}
	\label{fig:c_pi}
\end{figure}

The von Neumann entropy $S$ of a density matrix $\rho$, with eigenvalues $\{\lambda_i\}$, is defined by
\beq S=-\textrm{Tr}\,(\rho\log_2\,\rho)=-\sum_i \lambda_i\log_2\lambda_i\,, \eeq
and it constitutes a relevant measure of bipartite entanglement. The von Neumann entropy of the reduced density matrices of equal-sized bi-partitions of the Prime state, with an even number of qubits, was numerically calculated in Ref. \cite{entangl}, and found to scale linearly with the size of the bi-partition. 
To be precise, the best fit to a line, for $n=20\,$--$\,30$ qubits and for the natural equal-sized bi-partition, is $S(n)=0.88612902\,\frac{n}{2}-1.30405956$. This result  indicates that the Prime state is highly entangled, but not maximally so, because the {\sl maximal} possible scaling for the entropy would be linear in $\frac{n}{2}$ with a slope equal to 1.
Random states, i.e. those with random complex coefficients, follow a ``volume law'', with an entanglement entropy of half chain that scales as $n/2-1/2$ for big $n$, where $n$ is the number of qubits \cite{DonPage}. If we restrict to real 	positive random coefficients, the scaling is $(1 - 2/\pi) \,n/2 = 0.363\dots n/2$ \cite{GroverFisher}. So the Prime state is not a typical random positive state either.
Notice as well that the high entanglement of the Prime state implies that it is not possible to apply standard classical techniques such as Matrix Product States (MPS), or other more-refined tensor networks, to efficiently simulate the Prime state on a classical computer \cite{JI}.

\begin{figure}[t!]
	\centering	
	\includegraphics[scale=0.6]{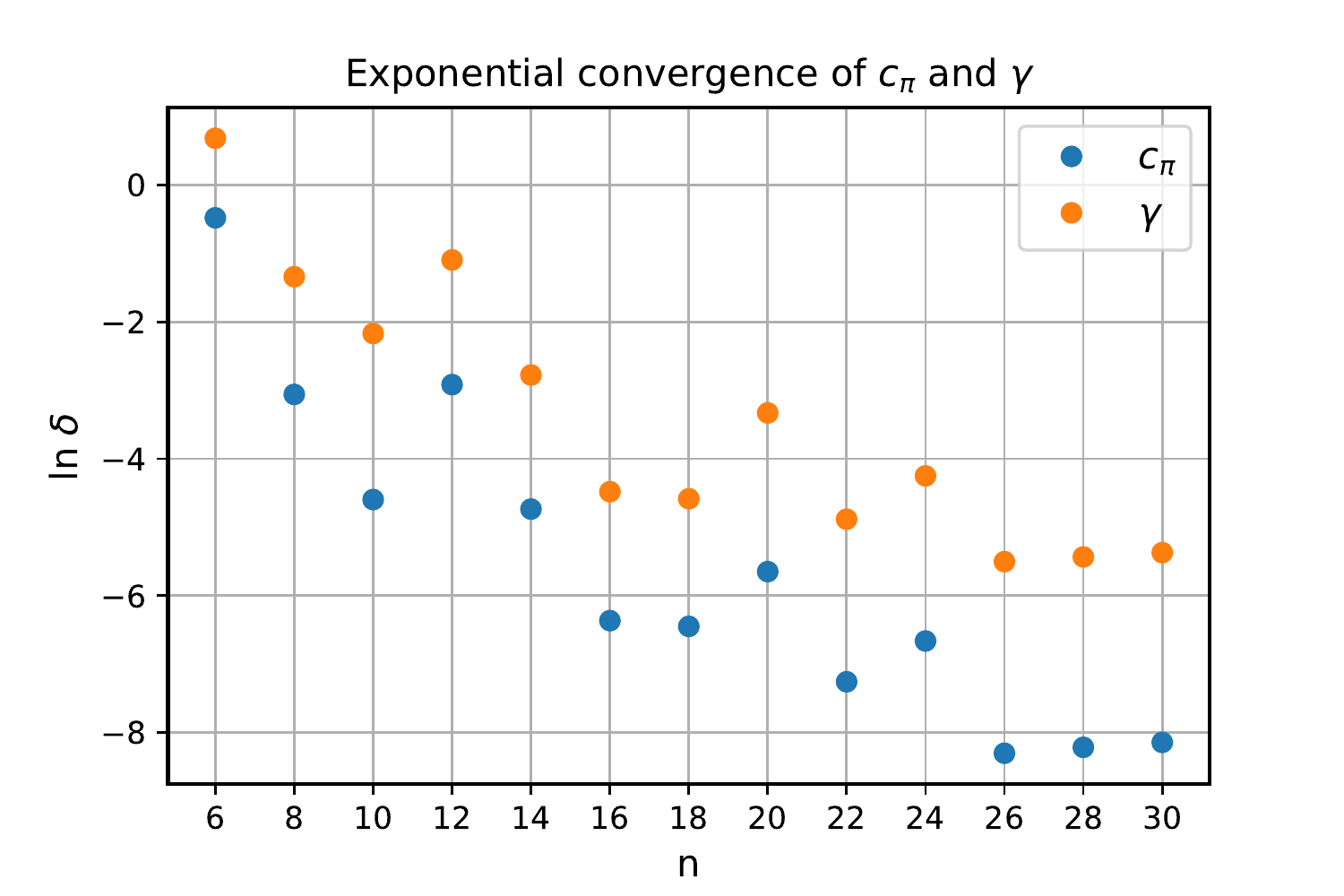}
	
	\caption{\small{Natural logarithm of the absolute value of the difference $\delta$, between the observed and predicted values of $c_\pi$ and $\gamma$, up to $n=30$ qubits. That is, $\delta\equiv \left|c_\pi(n)-H\left(\frac{3}{\pi^2}\right)\right|$ and $\delta\equiv \left|\gamma(n)-\left(1+\frac{3}{\pi^2}\right)\right|$, respectively. An exponential decrease in $\delta$ is observed before reaching a plateau.}}
	\label{fig:exp}
\end{figure}
\bigskip

We herein conjecture that the entropy of the natural equal-sized bi-partition of the Prime state of $n$ qubits is asymptotically given by
\beq 
\label{3pi2} 
S(n)= c_\pi\,\frac{n}{2}-\gamma\quad,\quad c_\pi=H\left(\frac{3}{\pi^2}\right)\quad,\quad\gamma=1+\frac{3}{\pi^2} \,,\eeq
where $H(p)\equiv -p\log_2(p)-(1-p)\log_2(1-p)$ is the Shannon entropy and $3/\pi^2=1/(2\,\zeta(2))$ is equal to a half of  the asymptotic density of odd square-free integers \cite{NumberTheory}. The constant $3/\pi^2$ also appears in the study of topological dynamical systems. In particular, it has been shown that it gives half the topological entropy of the square-free flow \cite{S11,CS11}.
The term -1 in the intercept comes from the fact that the least relevant qubit is
basically in the state $|1\ra$, because all primes but 2 are odd, and hence this qubit does not contribute to the entropy in the asymptotic limit.
In order to test the conjectured Eq. \eqref{3pi2}, we have computed the entropy of the Prime state up to $n=30$ qubits. Diagonalization using quadruple-precision floating-point numbers was implemented to meet the precision demanded, for large values of $n$, by the observed oscillatory behaviour of the slope $c_\pi$. To the best of our knowledge, this is the first open-source library that diagonalizes matrices using float128; as a matter of fact, this library works in arbitrary precision. The code is publicly available on GitHub, together with all the numerical results presented in this paper \cite{github}.

\begin{figure}[t!]
	\centering
	\includegraphics[scale=0.57]{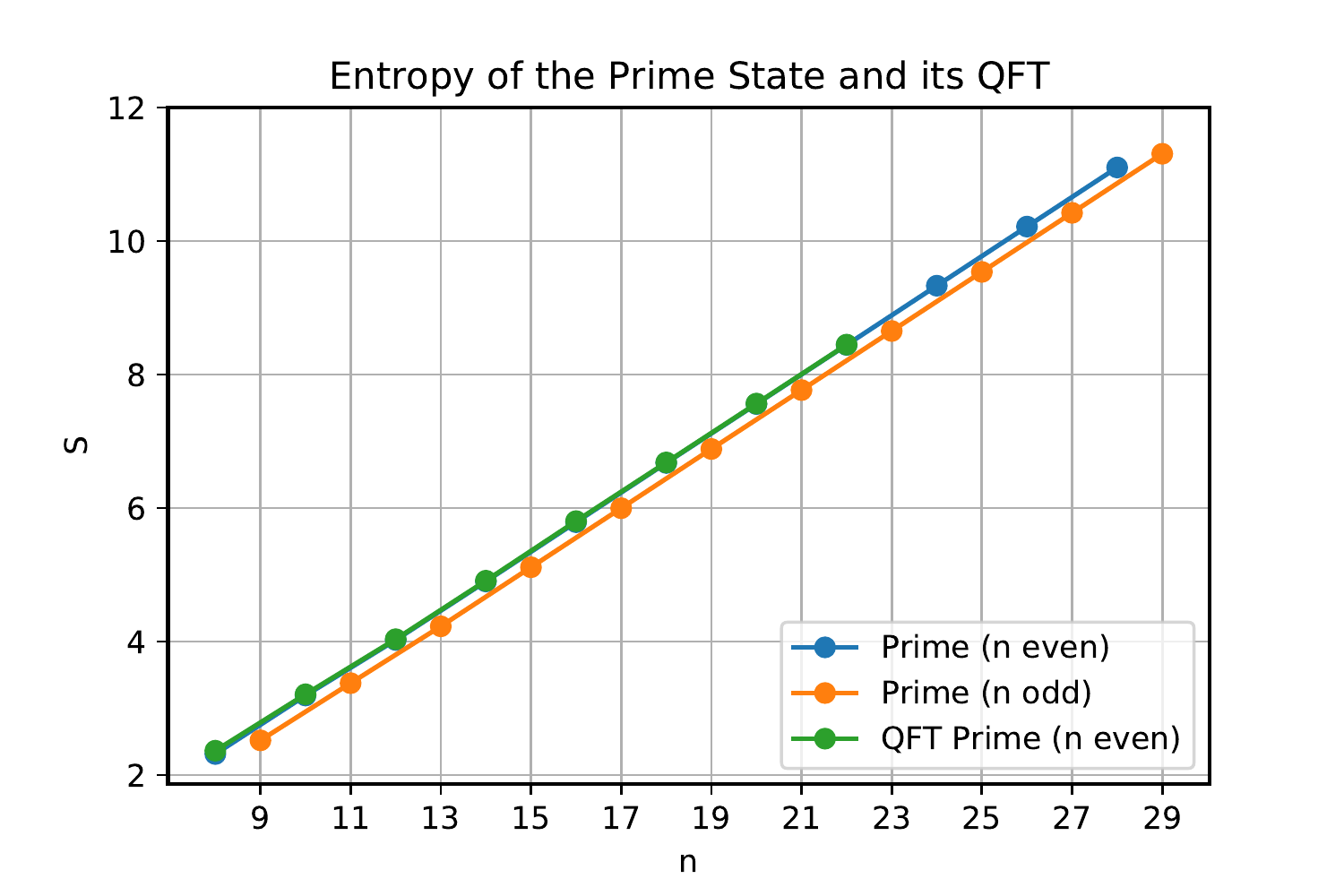}
	\includegraphics[scale=0.57]{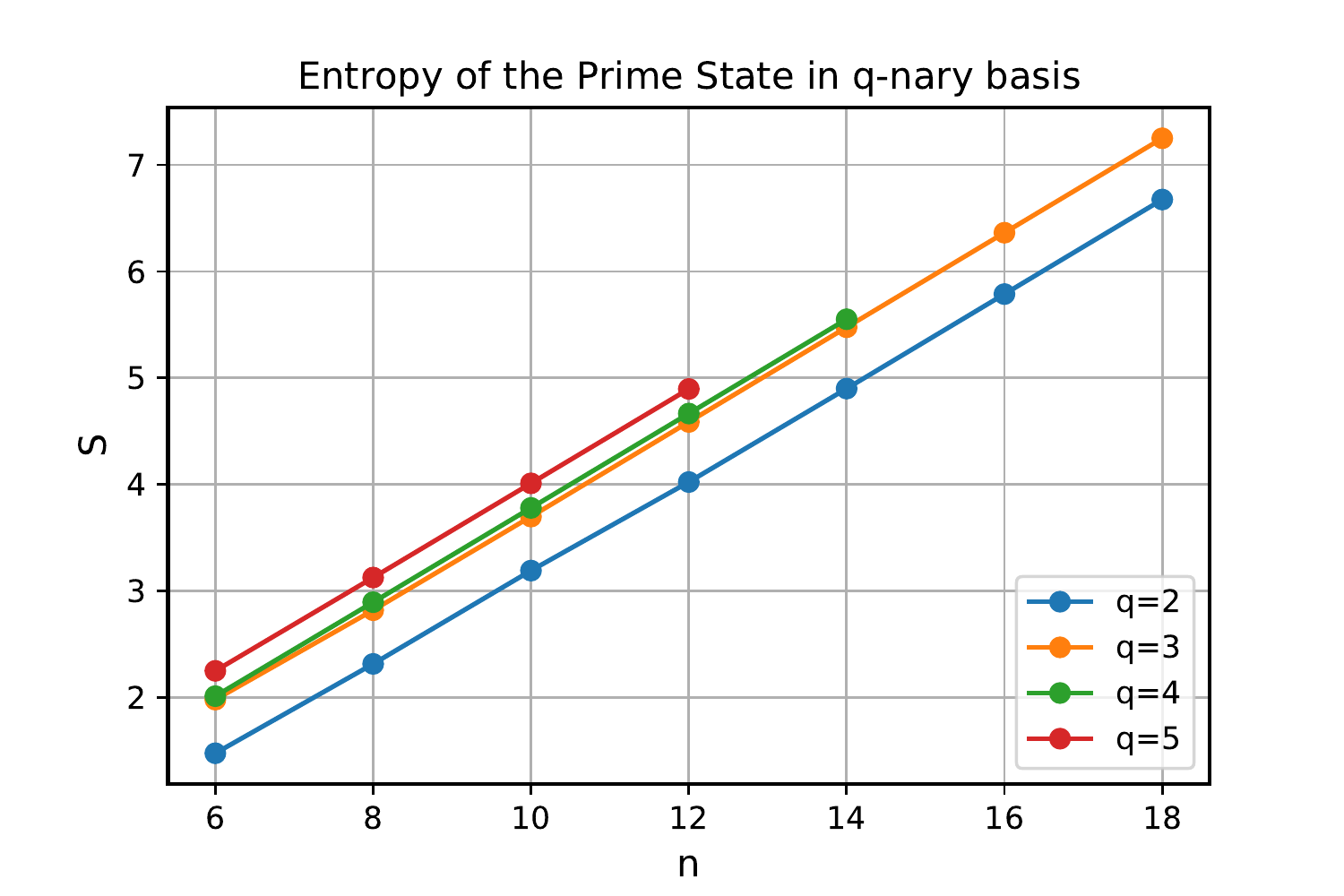}
	
	\caption{\small{top) Von Neumann entropy $S$ of the Prime state of $n$ qubits (up to $n=29$) and its Quantum Fourier Transform (up to $n=22$), for the natural bi-partition of size $\lfloor \frac{n}{2}\rfloor$. Notice that the entropies of the Prime state and its QFT are almost identical; bottom) Von Neumann entropy $S$ for the natural equal-sized bi-partition of Prime states expressed in $q$-nary bases using $n$ qudits.}}
	\label{fig:odd+bases}
\end{figure}
\bigskip

Figure \ref{fig:c_pi}-top shows the difference in entropy between two consecutive values of $S(n)$, for even $n$. That is, the dots in the plot correspond to 
\beq 
\label{c_pi} c_\pi(n)=S(n)-S(n-2)\,,\eeq
up to $n=30$ qubits. Oscillatory convergence towards the predicted value $H\left(\frac{3}{\pi^2}\right)=0.886082085\dots$ is observed, in good agreement with Eq. \eqref{3pi2}. The value of the intercept $\gamma$, obtained as
\beq \label{c'} \gamma(n)=\left(S(n)-S(n-2)\right)\frac{n}{2}-S(n)\,,\eeq
is shown in Fig. \ref{fig:c_pi}-bottom, up to $n=30$ qubits. Oscillatory convergence and good agreement with the predicted value $1+\frac{3}{\pi^2}=1.30396355\dots$ is observed as well. As a matter of fact, the behaviour of the convergence is almost identical in both cases, with an exponential approximation toward the conjectured values as the number of qubits grows, as shown in Fig. \ref{fig:exp}. The remarkably-similar fashion in which $c_\pi$ and $\gamma$ converge may be seen as an indication that both conjectures are closely related.

\begin{figure}[t!]
	\centering
	\includegraphics[scale=0.57]{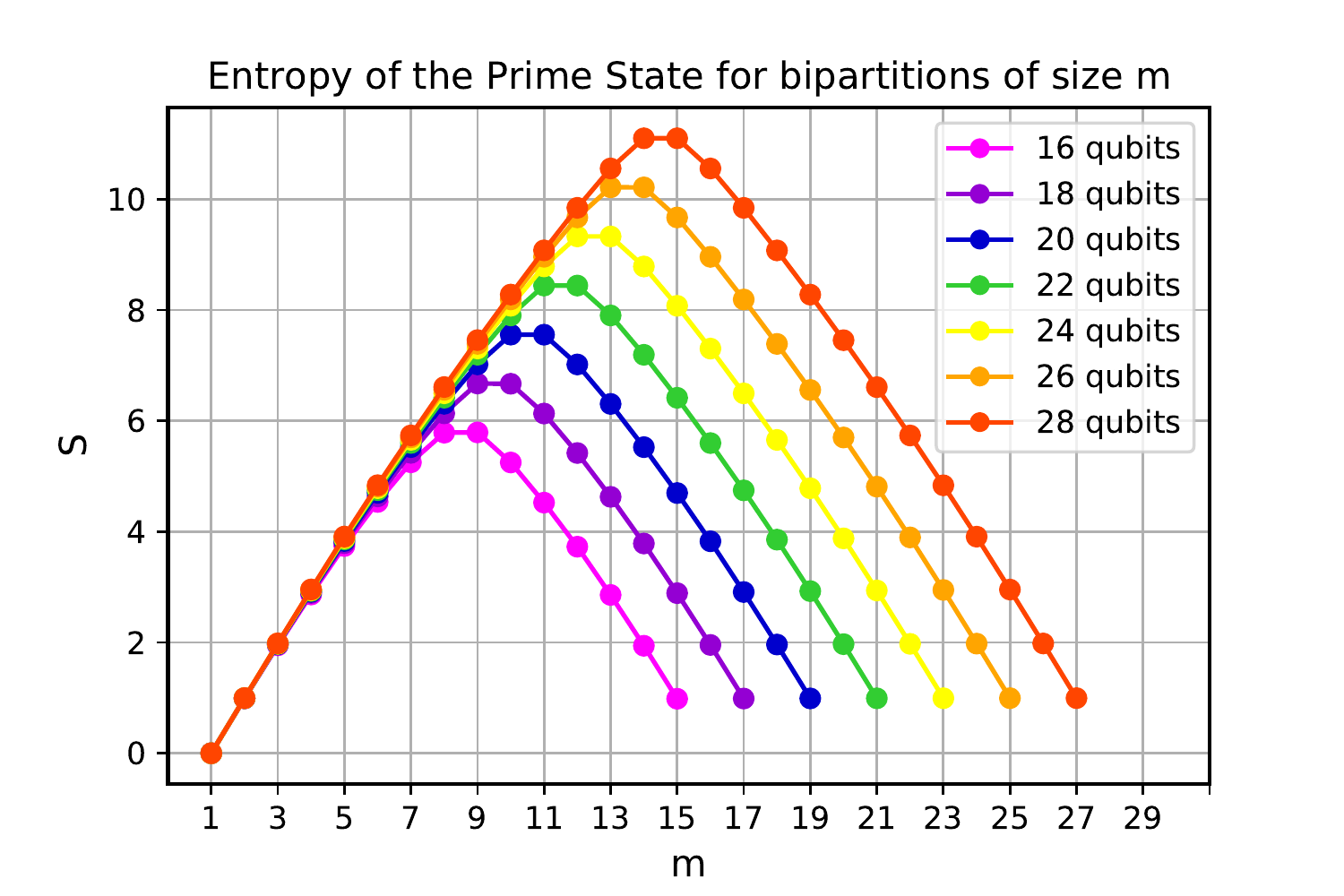}
	
	\includegraphics[scale=0.57]{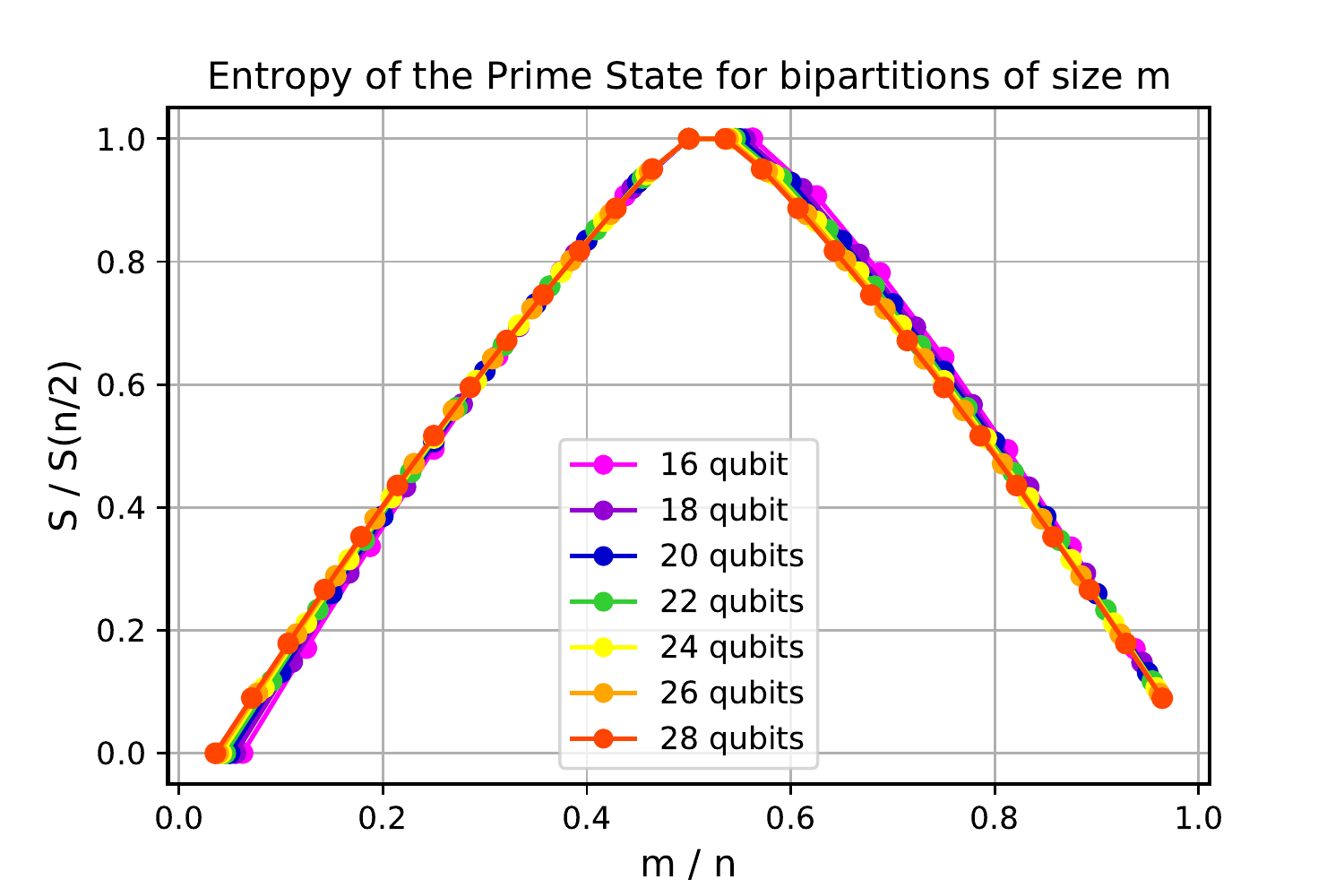}
	
	\caption{\small{top) Von Neumann entropy $S$ of Prime states with different number of qubits, up to $n=28$, for the natural bi-partition that separates the $m$ first (\ie least significant) qubits; bottom) Same von Neumann entropy $S$, rescaled by $S(n/2)$.}}
	\label{fig:M}
\end{figure}
\bigskip

We have also extended the computations to the natural equal-sized bi-partition of size $\lfloor \frac{n}{2}\rfloor$, with odd $n$ up to $n=29$ qubits, see Fig. \ref{fig:odd+bases}-top. The scaling of the entropy is very similar in this case to that of the Prime state with an even number of qubits, as it could be expected. Besides,
we have computed the von Neumann entropy of Prime states for qudit systems, with local dimensions $d=3,4,5$, for natural equal-sized bi-partitions, see Fig. \ref{fig:odd+bases}-bottom and Table \ref{tab:states}. This is equivalent to expanding the  prime numbers in the ternary basis, quaternary basis, etc. and putting them in superposition. We find that the scaling is independent of the basis, \ie linear with slope $\sim 0.88\dots$, so in this respect it is universal.
This result should be expected, because otherwise there would exist in some sense a `preferred' basis for expressing  prime numbers.
It is also truly noteworthy that we have found the same scaling (and almost the same values) for the Von Neumann entropy of the natural equal-sized bi-partitions of the QFT of the Prime state, see Fig. \ref{fig:odd+bases}-top and Table \ref{tab:states}.

Furthermore, we have computed the entropy for natural bi-partitions of size $m$ of the Prime state, for different number of qubits $n$, up to $n=28$ qubits, see Fig. \ref{fig:M}-top. Assuming a scaling behaviour  of the form $S(m,n)=n\,f(m/n)$, it follows that 
%
\beq \frac{S(m,n)}{S(n/2,n)}=\frac{f(m/n)}{f(1/2)}\,. \label{m/n}\eeq
This is precisely the behaviour observed when plotting the l.h.s of Eq. \eqref{m/n} against $m/n$, see Fig. \ref{fig:M}-bottom. This realization allows for a numerical approximation to $f(m/n)$ as a Fourier series, see Appendix D for the precise values of the first few coefficients. It follows that an estimate of $S(m,n)$ may be obtained from this series for {\sl any} value of $n,m$.

\begin{table}[t!] 
    \centering
	\begin{tabular}{|l|l|c|} \hline
		
		$State$ & $Slope$ & $Intercept$ \\\hline
		Prime (qubits)&    0.885791   &  -1.299313 \\\hline
		Prime (qutrits) & 0.887102 & -0.733640  \\\hline
		Prime (quatrits) &  0.885822 &  -0.649876 \\\hline
		Prime (5-qudits) &  0.886266 & -0.421618  \\\hline
		QFT Prime & 0.882952 & -0.382451\\\hline
		Arithmetic Prime (mod 4) & 0.884033  & -1.885681  \\\hline
		Arithmetic Prime (mod 8) &  0.883698& -2.575452  \\\hline
		Arithmetic Prime (mod 16) &  0.884932  & -3.339827  \\\hline
		Arithmetic Prime (mod 32) & 0.888128   & -4.169144  \\\hline
		M\"obius & 0.999843   & -0.953885 \\\hline
		Starry Prime & 0.985580   &  -0.948598 \\\hline
		
	\end{tabular}
	\caption{\small{Characterization of the linear scaling of the von Neumann entropy, for several number-theoretical quantum states. The entropy corresponds to the equal-sized natural bi-partition, and the slopes and intercepts have been calculated using Eq. \eqref{c_pi} and Eq. \eqref{c'}, for $n=30$; except for the M\"obius state ($n=28$), the QFT of the Prime state ($n=22$), and for qudit systems ($n=18,14,12$ for $d=3,4,5$ respectively). In the case of arithmetic Prime states, averages over the different arithmetic progressions are shown.}}
	\label{tab:states}
\end{table}

\subsection{Arithmetic Prime states}

\begin{figure}[t!]
	\centering
	\includegraphics[scale=0.6]{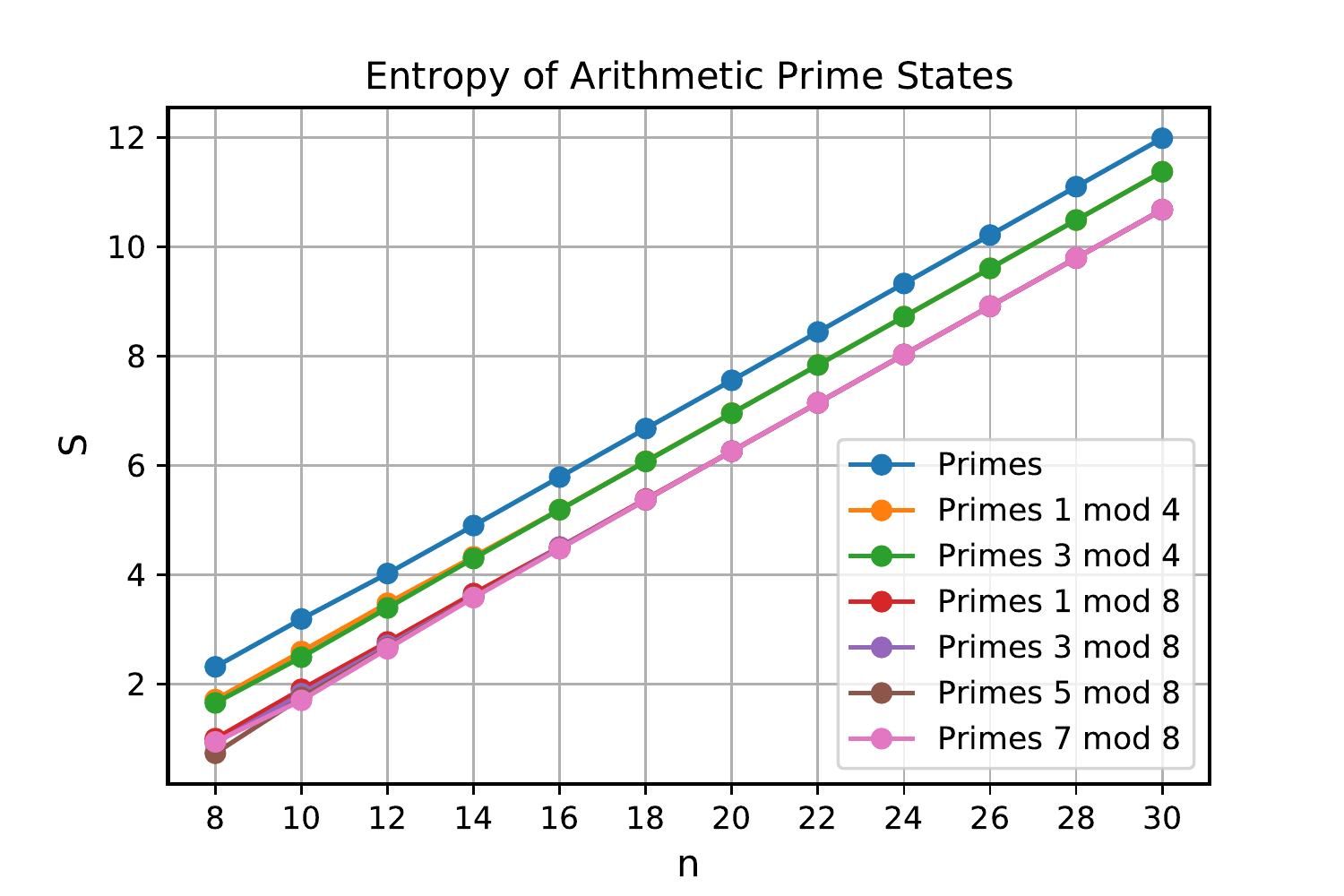}
	
	\caption{\small{Von Neumann entropy $S$ for the natural equal-sized bi-partition of Prime states in arithmetic sequences, compared to that of the Prime state, up to $n=30$ qubits. Observe the collapse of data for the sequences with the same moding,  when $n$ is large. }}
	\label{fig:mod4}
\end{figure}
\bigskip

Other number-theoretical quantum states apart from the Prime state will be defined below, and their entanglement traits studied. Table \ref{tab:states} summarizes our findings.
We define arithmetic Prime states as the uniform superpositions of primes belonging to arithmetic progressions, \ie
\beq |\mathbb{P}_{n;\,\alpha,\beta}\ra\equiv \frac{1}{\sqrt{\pi_{\alpha,\beta}(2^n)}} \sum_{\substack{p:\,prime\\p=\beta\,\textrm{mod} \,\alpha}}^{2^n} |p\ra \,,\eeq
with $\alpha,\beta$ coprime numbers. 
When $\alpha$ is a power of 2, a completely analogous derivation to that of Eq. \eqref{densmatrix} leads to an asymptotic expression for the reduced density matrix of arithmetic Prime states upon taking a natural bi-partition $A$--$B$ with the first (\ie least significant) $m$ qubits, 
\beq \label{arithdensmatrix} \overline{\rho}_A(\alpha) = \frac{\phi(\alpha)}{d}\left(\mathbbm{1}+l_N {\bf C}_m(\alpha)\right)\,,\eeq
where again $d=2^{m-1}$, $l_N\equiv\frac{Li_2(N)}{Li(N)}\xrightarrow{N\rightarrow \infty}\frac{1}{n\ln\,2}$, and ${\bf C}_m(\alpha)$ is a symmetric Toeplitz matrix defined by
\beq \label{HLalphaconstants} ({\bf C}_m(\alpha))_{i,j}\equiv (1-\delta_{i,j})\,C(\alpha\,|i-j|)\,.\eeq
Notice that Eq. \eqref{arithdensmatrix} generalizes the asymptotic expression of the reduced density matrix of the Prime state. Indeed, all primes but 2, which does not contribute in the asymptotic limit, belong to the arithmetic progression $2k+1$. Thus, $\phi(2)=1$, ${\bf C}_m(2)={\bf C}_m$, and we recover Eq. \eqref{densmatrix}. The validity of Eq. \eqref{arithdensmatrix} and Eq. \eqref{densmatrix} is conditional to the first Hardy-Littlewood conjecture and to the following asymptotic behaviour for the prime-number function $\pi_{\alpha;\,\beta,\beta'}(x)$,
\beq \label{alphabeta} \pi_{\alpha;\,\beta,\beta'}(x)\,\xrightarrow{\;x\rightarrow\infty\;}\,\frac{C(|\beta-\beta'|)}{\phi(\alpha)}\,Li_2(x)\,,\eeq
where $\pi_{\alpha;\,\beta,\beta'}$ is the number of prime  pairs $(p,p')$ such that $p,p'$ are primes smaller than, or equal to, $x$ of the form $p=\alpha k+\beta$, $p=\alpha k+\beta'$, with $(\alpha,\beta)$ and $(\alpha,\beta')$  coprimes pairs. This asymptotic behaviour, Eq. \eqref{alphabeta}, was numerically observed in Ref. \cite{entangl}.

\begin{figure}[t!]
	\centering
	\includegraphics[scale=0.6]{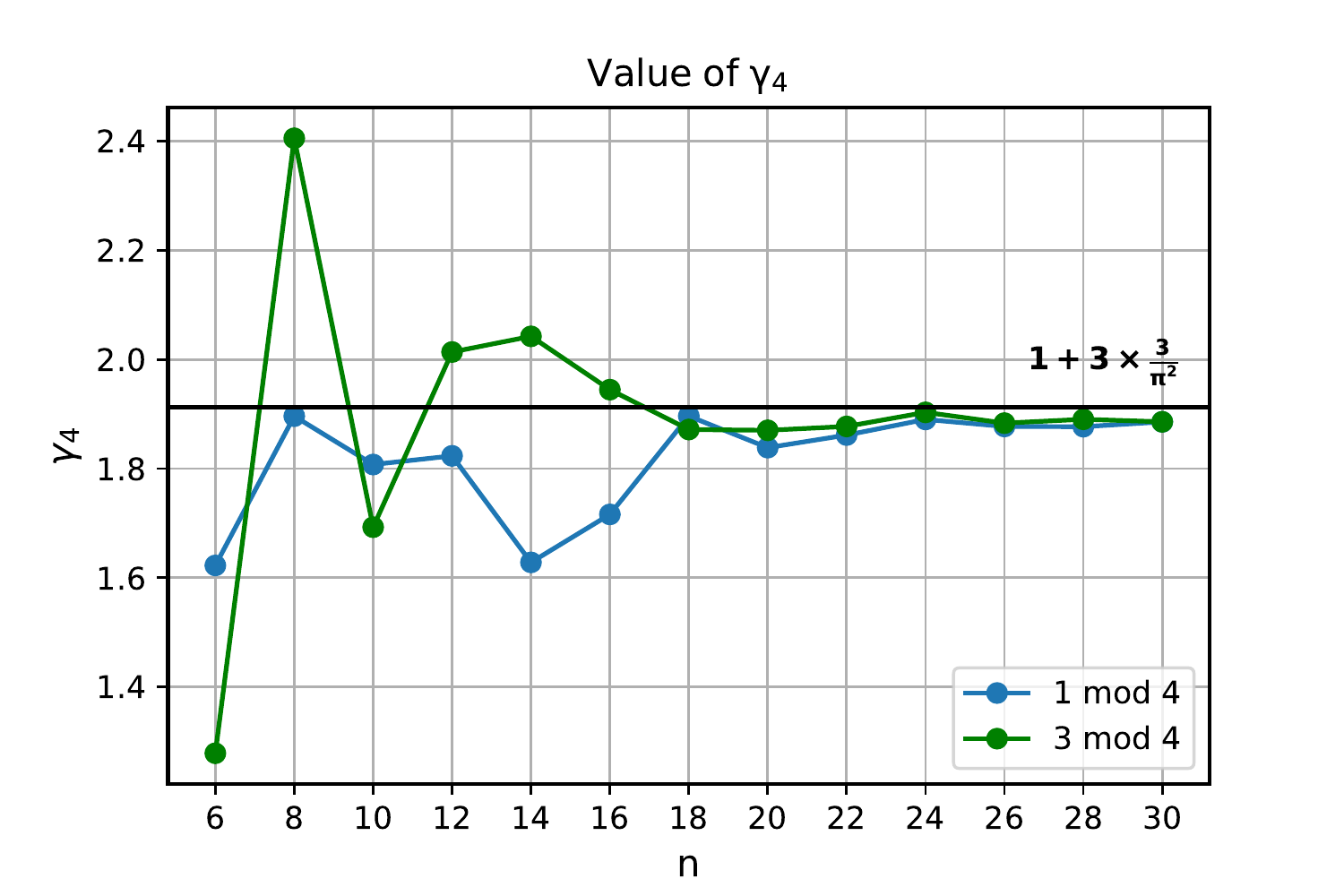}
	\includegraphics[scale=0.6]{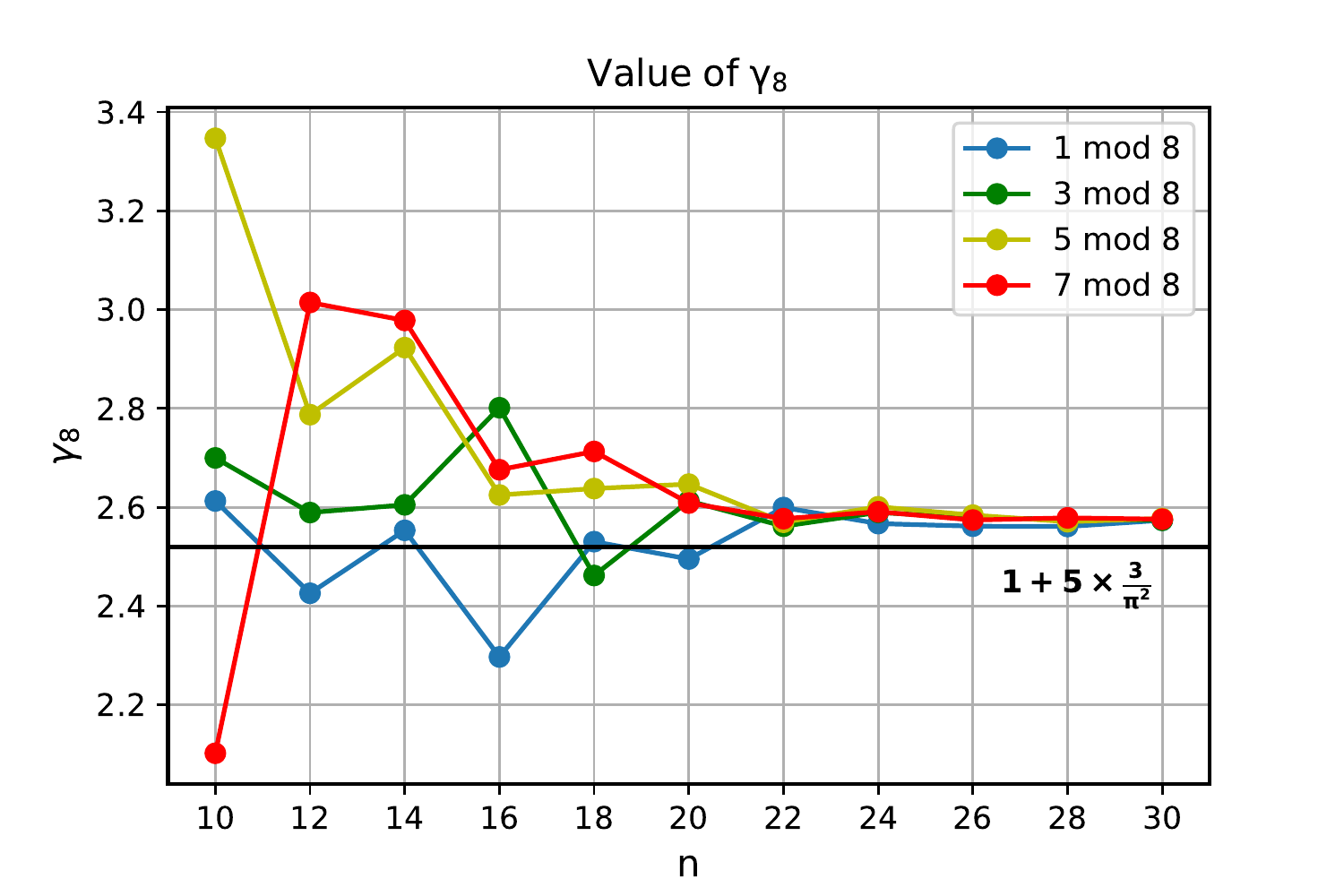}
	
	\caption{\small{ top) Value of $\gamma_4(n)$ computed according to Eq. \eqref{c'}, up to $n=30$ qubits}, compared to the value $1+3\times\frac{3}{\pi^2}$ in Eq. \eqref{gamma}; bottom) Value of $\gamma_8(n)$ computed according to Eq. \eqref{c'}, up to $n=30$ qubits, compared to the value $1+5\times\frac{3}{\pi^2}$ in Eq. \eqref{gamma}.}
	\label{fig:gamma_mod4}
\end{figure}

We have found that the scaling of the entropy of arithmetic Prime states, $|\mathbb{P}_{n;\,\alpha,\beta}\ra$, when $\alpha$ is a power of 2, is once again dictated by a straight line with slope $\sim0.884$, see Fig. \ref{fig:mod4} and Table \ref{tab:states}. The intercept, however, is different for distinct values of $\alpha$. It is also independent of $\beta$. The latter fact is easily understood from the asymptotic expression of the reduced density matrix $\overline{\rho}_A(\alpha)$, which does not depend on $\beta$. This is ultimately so because prime numbers are equally distributed on average among the different arithmetic progressions, for a given $\alpha$, as expressed in Eq. \eqref{pi,ab}. Hence, we label this constant as $\gamma_\alpha$. We have found the following behaviour for $\gamma_\alpha$ when $\alpha$ is a power of 2,
\beq \label{gamma} \gamma_\alpha \simeq 1 + s\frac{3}{\pi^2}\quad,\quad \alpha=2^k \quad,\quad s=1,3,5,\dots\,,\eeq
where $s$ is the $k$-th odd square-free number. Figure \ref{fig:gamma_mod4} shows the independence of $\gamma_4$ and $\gamma_8$ from $\beta$, for a large number of qubits, and the convergence towards a value close to that in Eq. \eqref{gamma}. Using the results obtained for $\gamma_4$, $\gamma_8$, $\gamma_{16}$ and $\gamma_{32}$ for $n=30$ qubits, averaged among the different progressions, the value $s=\left(\gamma_\alpha-1\right)\frac{\pi^2}{3}$ is found to be $s=2.9137\dots$, $s=5.1830\dots$, $s=7.6977\dots$ and $s=10.4260\dots$, for $\alpha=4,8,16,32$ respectively.

Taking into account all the results presented in this section, we generalize the conjecture in Eq. \eqref{3pi2}, stating that, for Prime and arithmetic Prime states of $n$ qudits, the von Neumann entropy $S$ of equal-sized bi-partitions is asymptotically given by
\beq S(n)= c_\pi\,\lfloor \frac{n}{2}\rfloor-\gamma_{\alpha;\,d} \quad,\quad c_\pi=H\left(\frac{3}{\pi^2}\right)\,,\eeq
where $\lfloor x\rfloor$ is the floor function and $\gamma_{\alpha;\,d}$ is a constant that depends upon the specific arithmetic progression used to define the state, as well as the qudit basis in which the numbers are expressed; in addition, it also depends on whether $n$ is even or odd.

\bigskip

\subsection{Quantum relatives of the Prime state}

\begin{figure}[t!]
	\centering
	\includegraphics[scale=0.6]{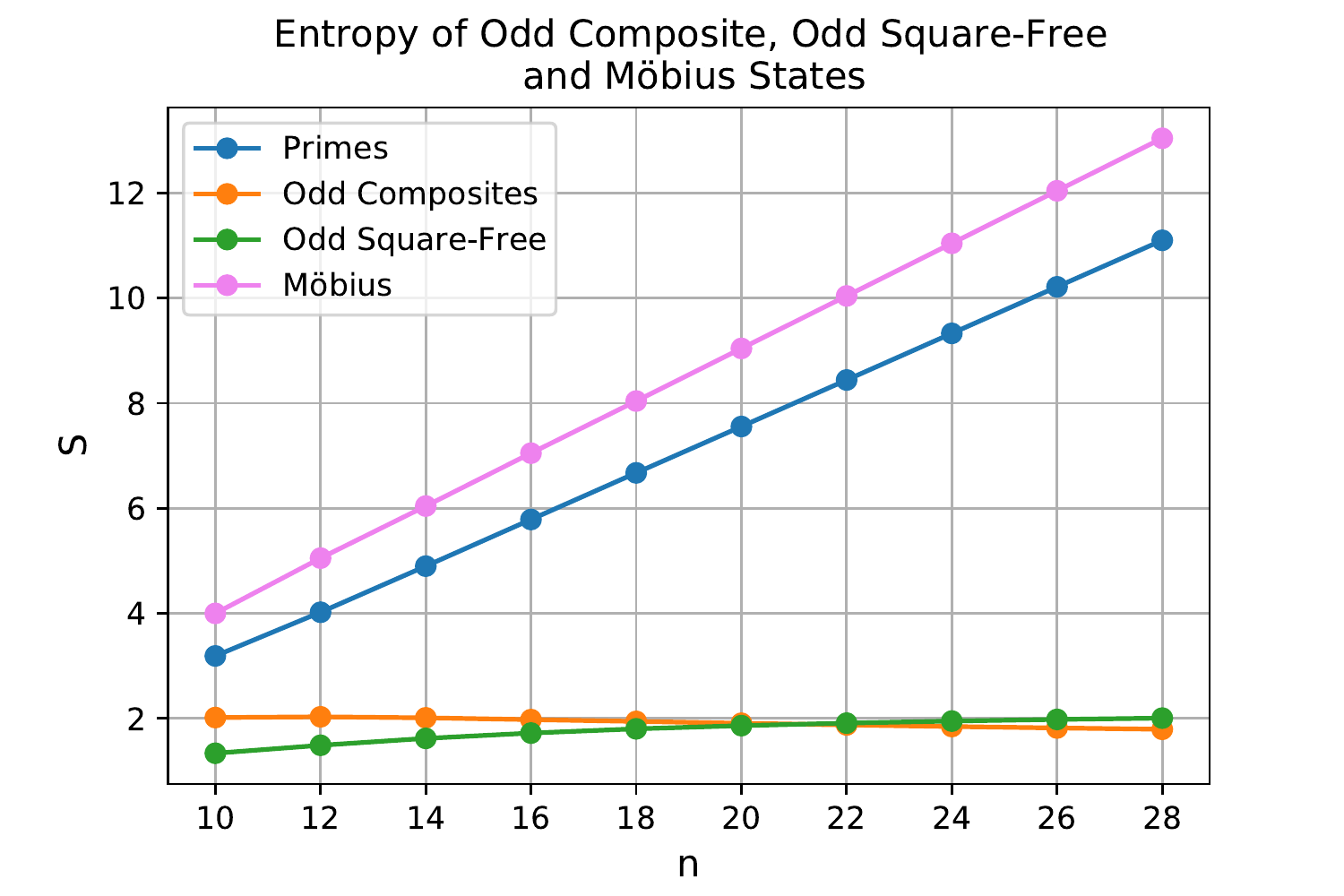}
	
	\caption{\small{Von Neumann entropy $S$ of the natural equal-sized bi-partition of the odd composite, odd square-free and M\"obius states, up to $n=28$ qubits.}}
	\label{fig:composite}
\end{figure}

One may ask whether it could be possible to construct a state that is the uniform superposition of odd composite (\ie non-prime) numbers, and whether or not this state holds large entanglement, in the  hope of  finding  a shortcut to the distribution of primes by looking at the distribution of composites. This state is defined by
\beq |\slashed{\mathbb{P}}_n\ra \equiv \frac{1}{\sqrt{C_{c,n}}} \sum_{\substack{c:\,odd\\c:\,composite}}^{2^n-1} |c\ra\quad,\quad C_{c,n}=2^{n-1}-\pi(2^n)\,.\eeq
We have computed the von Neumann entropy for the natural equal-sized bi-partition of such odd composite state, up to $n=28$ qubits, and found that this entropy remains approximately constant around a mean value of $\sim2$, irrespective of the number of qubits $n$, see Fig. \ref{fig:composite}. This behaviour is reminiscent of finitely-correlated systems away from criticality \cite{constant}. A slight decline in the entropy can be observed as $n$ increases, though. An asymptotic expression for the reduced density matrix of the odd composite state upon taking a bi-partition $A$-$B$ with the first (\ie least significant) $m$ qubits, can be obtained, as shown in Appendix E. It is given, when $n,m\rightarrow\infty$, by
\beq \label{composite1} \overline{\rho}_A = \frac{1}{d} \left(\mathbbm{1}+{\bf P}_m \right)\,, \eeq
where
\beq \label{composite}
	 \left({\bf P}_m\right)_{ij}= \left\{ \begin{array}{lc}  0 & {\rm if}\; i=j \\ 1 & {\rm if}\; i\neq j
\end{array}\right. \,.\eeq
From Eq. \eqref{composite1}, the von Neumann entropy $S$ in the asymptotic limit is given by
\beq S\simeq 0 \,,\eeq
in agreement with the observed numerical results.  One might hope that, because the entanglement entropy of the odd composite state is small, there could indeed be a shortcut to efficiently compute the prime number distribution by using tensor networks. But this is
incorrect, as can be argued readily. Most numbers are composite, and the composite state carries their equal superposition. This very high density of equally-weighted  superpositions  approaches the product state $|\phi\ra=\frac{1}{\sqrt{2^n}}\,\sum_{i=0}^{2^n-1}|e_i\ra$. Thus, tensor networks of low ancillary dimension only describe
trivial properties of the composite state. To learn something about primes from the composite state, we need to go to the subleading terms in any figure of merit, which implies exponential effort in the tensor network description.

Another number-theoretical quantum state, relative of the Prime state, is the odd square-free state, \ie the uniform superposition of odd square-free numbers, which are odd integers that do not contain any prime raised to a power larger than 1 in its prime decomposition. This state is defined by
\beq \label{free} |\mathbb{S}_n \ra \equiv \frac{1}{\sqrt{C_{s,n}}}\,\sum_{s:\, odd}^{2^n} \left|\mu(s)\right|\, |s\ra\quad,\quad C_{s,n}=\sum_{s:\,odd}^{2^n} |\mu(s)|\,. \eeq
We have computed the von Neumann entropy of this state, up to $n=28$ qubits, and found that it follows a similar behaviour to that of the odd composite state, although now the entropy shows a slight increase as the number of qubit grows (see Fig. \ref{fig:composite}). The crossing point of both entropies is located at $n\sim20$ qubits.

\bigskip

Another possibility is to define a uniform superposition of square-free numbers and let the signs of the amplitudes be given by the M\"obius function. This state, that we shall call M\"obius state \cite{entangl}, is then defined by
\beq |\mu_n\ra \equiv\frac{1}{\sqrt{C_{\mu,n}}}\, \sum_{s=1}^{2^n} \mu(s) \,|s\ra\quad,\quad C_{\mu,n}=\sum_{s=1}^{2^n} \left|\mu(s)\right|\,.\eeq
The von Neumann entropy of the natural equal-sized bi-partition of the M\"obius state scales linearly with a slope close to 1. This means that the M\"obius state is maximally entangled, up to a constant (see Fig. \ref{fig:composite} and Table \ref{tab:states}).
Note that the appearance of negative signs in the coefficients of the superposition allows for the appearance of almost maximal entanglement. 
This shows that there are two ways of getting maximal entanglement. One is to play with random coefficients in the quantum state. Natural cancellations make the reduced density matrix very mixed. Alternatively, we may impose that all coefficients are zero or one (up to normalization). Then, maximal entanglement is related to the thin distribution of numbers present in the state.

\begin{figure}[t!]
	\centering
	\includegraphics[scale=0.6]{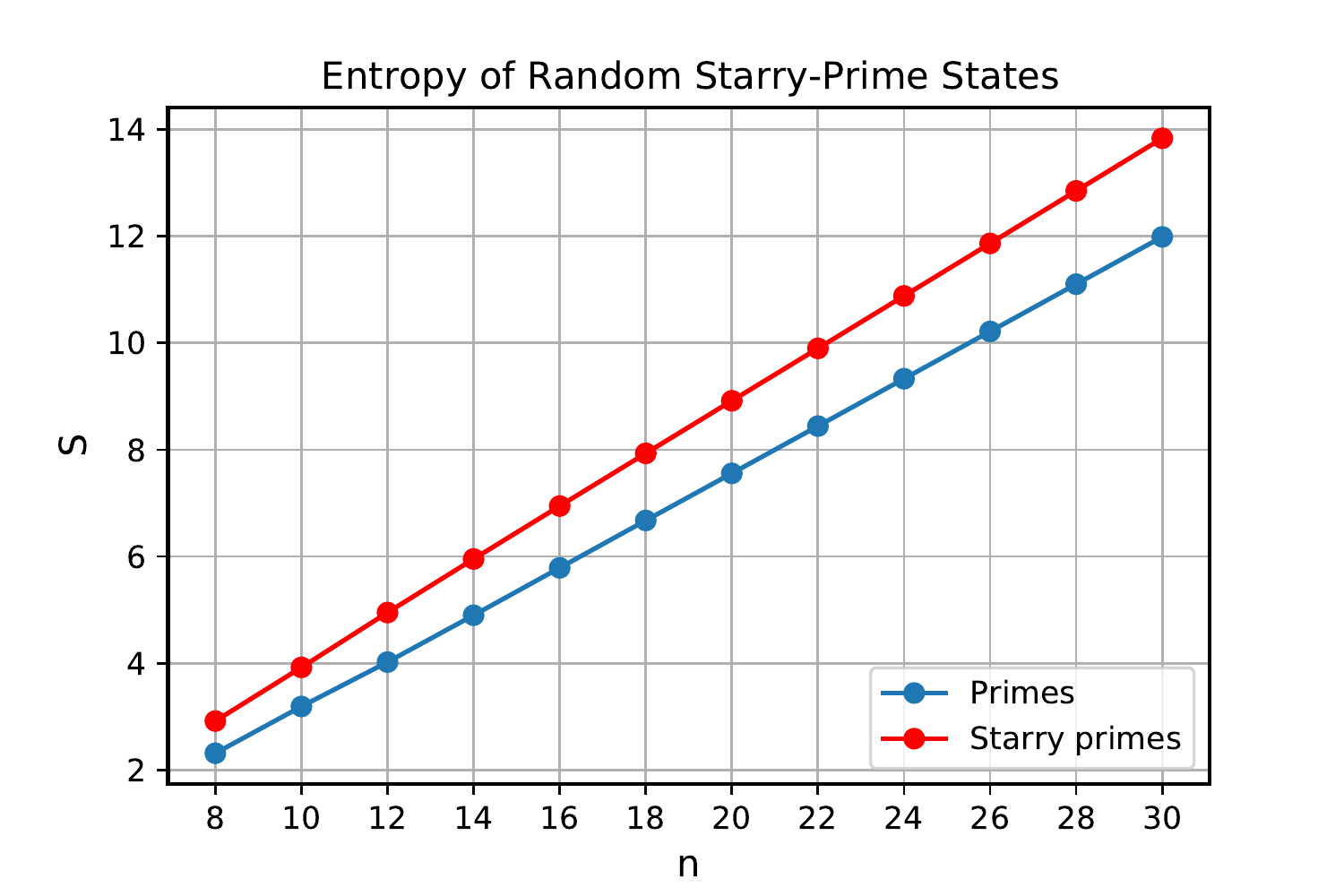}
	
	\caption{\small{Von Neumann entropy $S$ of random starry-Prime states for the natural equal-sized bi-partition, compared to that of the Prime state, up to $n=30$ qubits. Each red dot is an average over 10 random starry-Prime states.}}
	\label{fig:fakes}
\end{figure}

Finally, we have computed the entropy of the natural equal-sized bi-partition of {\em starry}$\,$-Prime states. 
Starry primes are integer numbers, such that the $n$-th starry prime, $p^*_n$, satisfies
\beq \label{fakes} p_n\leq p^*_n\leq p_{n+1}\,, \eeq
where $p_n$ is the $n$-th prime. These numbers are remarkable in that a $\zeta^*$ function, defined via the Euler product formula, \ie
\beq \zeta^*(s)\equiv \prod_{p^*}\frac{1}{1-{p^*}^{-s}} \,,\eeq
has an analytic extension with exactly the same zeros, and the same multiplicities, than the original Riemann zeta function \cite{fakes}. 
The starry-Prime state is then defined by
\beq \label{fakestate} |\mathbb{P^*}_n\ra\equiv \frac{1}{\sqrt{\pi(2^n)}}\,\sum_{\substack{p^*:\,starry\\\quad prime}}^{2^n-1} |p^*\ra\,,\eeq
where $\{p^*\}$ is any set of starry primes up to $2^n-1$. We have found that the entropy of the natural equal-sized bi-partition of random starry-prime states --where the terms entering into the superposition Eq. \eqref{fakestate} were taken at random according to Eq. \eqref{fakes}--, scales almost in a maximal way, \ie linearly in $\frac{n}{2}$ with slope $\sim1$ (see Fig. \ref{fig:fakes} and Table \ref{tab:states}). This result shows that primes are less random than starry primes. The extra element of randomness in the choice of numbers within intervals defined by primes is sufficient to raise the slope of the entanglement entropy to a value near 1.

\bigskip

\section{Conclusions}

Prime numbers are classical  mathematical objects.
However, using a quantum computer, they can be efficiently placed in a linear superposition,   the Prime state.
This cast of a classical sequence  onto a quantum state brings a novel quantum perspective to Number Theory that we are just  beginning to explore.

In this work, we have shown how Chebyshev-like biases arise from 
the Quantum Fourier Transform of the Prime state, and how to estimate them from measurements.
We have also delved into the entanglement properties of the Prime state, 
which turn out to be intimately related to the correlations between pairs of prime numbers. 
These correlations were postulated by Hardy and Littlewood almost a century ago, and  played a critical role in the connection between Number Theory 
and the Theory of Quantum Chaos based on heuristic analogies. 
According to it,  prime numbers are classical objects 
(labelling periodic  orbits in a chaotic dynamical system),  
while the Riemann's zeros are quantum objects (eigenenergies of a hypothetical Riemann Hamiltonian). 
In our approach, the Hardy and Littlewood conjecture provides  us with the entanglement Hamiltonian 
of the Prime state, whose spectrum we have found to be characterized by the odd square free integers. 
Furthermore, the entanglement entropy of the Prime state is closely related to the Shannon entropy 
of half the density of the square-free integers. 
The previous  result also applies to the quantum superposition of prime numbers 
in arithmetic progressions, which is a kind of quantum version of Dirichlet's theorem about these sequences.
Likewise, we have verified that the entanglement entropy of the Prime state
does not depend on the basis used, \ie qubits, qutrits, etc., which is an expected result 
but that confirms its universality. We have furthermore characterized entanglement properties of other quantum states built upon different sequences of numbers.
From a more practical point of view, we have developed an open-source library that diagonalizes matrices using floats of arbitrary precision.

We expect this novel quantum approach to Number Theory to be fruitful in the future, as it may help us deepen into the profound mysteries of numbers.

\bigskip
\noindent{{\sl \bf Acknowledgements.} We thank J. Andrade, A. Botero, J. Keating, A. LeClair, G. Mussardo, D. Pérez-García and S. Ramos-Calderer for useful discussions. DGM and JIL acknowledge {\minibox{CaixaBank}} for its support of this work through Barcelona Supercomputing Center's project {\sl CaixaBank Computación  Cuántica}.  DGM and JIL are supported by Project PGC2018-095862-B-C22, and Project Quantum CAT (001- P-001644) co-funded by Generalitat de Catalunya and the European Union Regional Development Fund. SC is supported by the European Research Council under the European Unions Horizon 2020 research and innovation Programme (grant agreement number 740006). GS is supported by the Ministerio de Ciencia, Innovación y Universidades (grant PGC2018-095862-B-C21), the Comunidad de Madrid (grant QUITEMAD+ S2013/ICE-2801), the Centro de Excelencia Severo Ochoa Programme (grant SEV-2016-0597) and the CSIC Research Platform on Quantum Technologies PTI-001.

\newpage

\onecolumngrid
\newpage

\appendixpageoff
\appendixtitleoff
\renewcommand{\appendixtocname}{Supplementary material}
\begin{appendices}
  \section*{ Appendices}

\section{ Computation of relevant peaks in $|\hat{\mathbb{P}}_n\ra$}
\label{QFTpeaks}

\noindent We compute here the most relevant peaks of the QFT of the Prime state, $|\hat{\mathbb{P}}_n\ra$. We use Eq. \eqref{QFTdef}, where $x_j=\frac{1}{\sqrt{\pi(N)}}$ for prime $j$ and 0 otherwise. Recall that $N=2^n$ is the dimension of the Hilbert space.
\beq \begin{split} P(N/2)=\quad & \frac{1}{N\pi(N)} \left| \sum_{p:\,prime} e^{p\pi i}\right|^2=\; \frac{1}{N\pi(N)}\left| \sum_{\substack{p:\,prime\\p=0\,\textrm{mod2}}} e^{p\pi i}+ \sum_{\substack{p:\,prime\\p=1\,\textrm{mod2}}} e^{p\pi i}\right|^2= \\\\  &\frac{1}{N\pi(N)} \left|\pi_{2,0}(N)-\pi_{2,1}(N)  \right|^2=\; \frac{1}{N\pi(N)} \left| 1 -(\pi(N)-1) \right|^2 = \quad \frac{\pi(N)^2-4\pi(N)+4}{N\pi(N)} \end{split}\eeq
\beq \label{N/3} \begin{split}  P(N/3)= \quad &\frac{1}{N\pi(N)} \left| \sum_{p:\,prime} e^{\frac{2}{3}\pi i\, p} \right|^2 =\; \frac{1}{N\pi(N)}\left| \sum_{\substack{p:\,prime\\p=0\,\textrm{mod3}}} e^{\frac{2}{3}p\pi i}+ \sum_{\substack{p:\,prime\\p=1\,\textrm{mod3}}} e^{\frac{2}{3}p\pi i} + \sum_{\substack{p:\,prime\\p=2\,\textrm{mod3}}} e^{\frac{2}{3}p\pi i}\right|^2= \\\\ &\frac{1}{N\pi(N)} \left|\pi_{3,0}(N)+\pi_{3,1}(N)\,e^{\frac{2}{3}\pi i}+\pi_{3,2}(N)\,e^{\frac{4}{3}\pi i} \right|^2=\\\\ &\frac{1}{N\pi(N)} \left|1+\left(-\frac{1}{2}+i\frac{\sqrt{3}}{2}\right)\pi_{3,1}(N)+ \left(-\frac{1}{2}-i\frac{\sqrt{3}}{2}\right)\pi_{3,2}(N) \right|^2= \\\\ & \frac{1}{N\pi(N)} \left|1-\frac{1}{2}\left(\pi_{3,1}(N)+\pi_{3,2}(N)\right)-i\frac{\sqrt{3}}{2}\, \Delta_{3;2,1}(N) \right|^2=\\\\ & \frac{1}{N\pi(N)}\left(1+\frac{1}{4}\left(\pi_{3,1}(N)+\pi_{3,2}(N)\right)^2-\pi_{3,1}(N)-\pi_{3,2}(N)+\frac{3}{4}\Delta_{3;2,1}(N)^2 \right) =\\\\ & \frac{\Delta_{3;2,1}(N)^2+\pi_{3,1}(N)\pi_{3,2}(N) -\pi(N)+2}{N\pi(N)}\end{split}\eeq
\beq \begin{split} P(N/4)= \quad &\frac{1}{N\pi(N)} \left| \sum_{p:\,prime} e^{\frac{1}{2}\pi i\, p} \right|^2=\; \frac{1}{N\pi(N)} \left| \sum_{\substack{p:\,prime\\p=2\,\textrm{mod4}}} e^{\frac{1}{2}p\pi i}+ \sum_{\substack{p:\,prime\\p=1\,\textrm{mod4}}} e^{\frac{2}{4} p\pi i} + \sum_{\substack{p:\,prime\\p=3\,\textrm{mod4}}} e^{\frac{2}{4} p\pi i}\right|^2= \\\\ &\frac{1}{N\pi(N)} \left|-\pi_{4,0}(N) +  \pi_{4,1}(N)\,e^{\frac{1}{2}\pi i} + \pi_{4,3}(N)\,e^{\frac{3}{2}\pi i} \right|^2 = \; \frac{1}{N\pi(N)}\left|-1 +  i \left(\pi_{4,1}(N)- \pi_{4,3}(N)\right) \right|^2= \\\\ &\frac{1+\Delta(N)^2}{N\pi(N)}  \end{split}\eeq
\beq \label{N/6} \begin{split} P(N/6)= &\quad  \frac{1}{N\pi(N)} \left| \sum_{p:\,prime} e^{\frac{2}{6}\pi i\, p} \right|^2=\; \frac{1}{N\pi(N)} \left| \sum_{\substack{p:\,prime\\p=2\,\textrm{mod6}}} e^{\frac{1}{3}p\pi i}+\sum_{\substack{p:\,prime\\p=3\,\textrm{mod6}}} e^{\frac{1}{3}p\pi i}+ \sum_{\substack{p:\,prime\\p=1\,\textrm{mod6}}} e^{\frac{2}{6} p\pi i} + \sum_{\substack{p:\,prime\\p=5\,\textrm{mod6}}} e^{\frac{2}{6} p\pi i}\right|^2=\\\\ &\frac{1}{N\pi(N)} \left|\pi_{6,0}(N)\,e^{\frac{2}{3}\pi i}+\pi_{6,3}(N)e^{\pi i} +\pi_{6,1}(N)\,e^{\frac{1}{3}\pi i}+\pi_{6,5}(N)\,e^{\frac{5}{3}\pi i} \right|^2 =\\\\ &\frac{1}{N\pi(N)} \left|\left(-\frac{1}{2}+i\frac{\sqrt{3}}{2}\right)-1+\left(\frac{1}{2}+i\frac{\sqrt{3}}{2}\right)\pi_{6,1}(N)+\left(\frac{1}{2}-i\frac{\sqrt{3}}{2}\right)\pi_{6,5}(N)\right|^2=\\\\ &\frac{1}{N\pi(N)} \left|\frac{\pi_{6,1}(N)+\pi_{6,5}(N)-3}{2}-i\frac{\sqrt{3}\,\left(\Delta_{6;5,1}(N)-1\right)}{2} \right|^2 = \\\\ & \frac{1}{N\pi(N)} \left(\frac{9+\pi_{6,1}(N)^2+\pi_{6,5}(N)^2+2\,\pi_{6,1}(N)\pi_{6,5}(N)-6\,\pi_{6,1}(N)-6\,\pi_{6,5}(N)}{4} +\frac{3\,\Delta_{6;5,1}(N)^2+3-6\,\Delta_{6;5,1}(N)}{4}\right)=\\\\  &\frac{\Delta_{6;5,1}(N)^2+\pi_{6,1}(N)\pi_{6,5}(N)-3\,\pi_{6,5}(N)+3}{N\pi(N)}  \end{split}\eeq

Given that $N/3$ and $N/6$ are not integers when $N=2^n$, ancillary qubits are required to represent rational numbers as binary fractions. A numerical check of the error incurred in the estimation of the peak when using nine ancillary qubits to represent the non-integer part of $N/3$ or $N/6$ is shown in Fig. \ref{fig:peak_error}, for $n=10$-30 qubits. The relative error is around $10^{-6}$ for $n=30$ qubits. If more precision is required, one can simply add more ancillas.

\begin{figure}[b]
    \centering
    \includegraphics[scale=0.6]{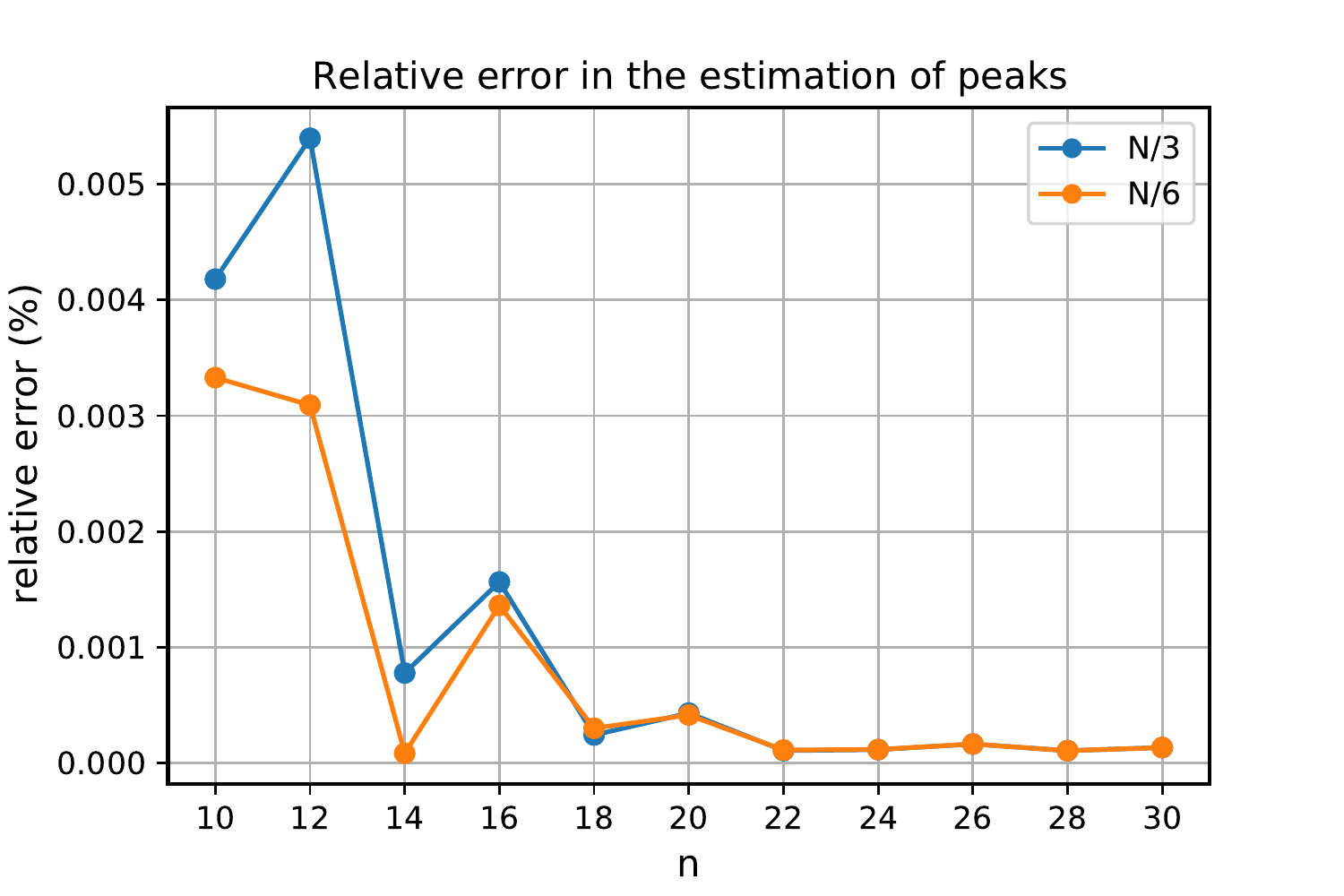}
    \caption{Relative percentage error in the estimation of Eqs. \eqref{N/3}, \eqref{N/6} due to the limited precision in the representation of $N/3$ and $N/6$, using nine ancillary qubits, for $n=10-30$ qubits.}
    \label{fig:peak_error}
\end{figure}

\section{Quantum Fourier Transform of a uniform superposition}
\label{app:QFTuniform}

\noindent Let us consider the definition of the QFT of a general quantum state $\ket$ of $n$ qubits, expressed in the computational basis, $\{|e_j\}$,
\beq \label{qft} \ket =\sum_{j=0}^{2^n-1} x_j\,|e_j\ra \xrightarrow{\quad QFT\quad} \sum_{k=0}^{2^n-1} y_k\,|e_k\ra\,,\eeq
where
\beq \label{y_k} y_k\equiv \frac{1}{\sqrt{2^n}}\,\sum_{j=0}^{2^n-1} x_j\,e^{2\pi i\,jk/2^n}\,,\eeq
and the $i$ in the exponent is the imaginary unit. The probability of measuring the $|0\dots0\ra$ state in Eq. (\ref{qft}) in the computational basis thus is:
\beq P(0)=|y_0|^{\,2}=\left|\frac{1}{\sqrt{2^n}}\,\sum_{j=0}^{2^n-1} x_j\,e^{0}\,\right|^2=\frac{1}{\,2^n\,} \,\left|\sum_{j=0}^{2^n-1} x_j\,\right|^2\,.\eeq 
If the state $\ket$ is a uniform superposition of $M$ computational-basis vectors, \ie $x_j=1/\sqrt{M}$ for all $j$ such that $x_j\neq0$, and $N=2^n$ is the dimension of the Hilbert space, then
\beq \label{counting} P(0)=\frac{1}{\,N\,}\,\left|\sum_{j=0}^{M-1} \frac{1}{\sqrt{M}}\,\right|^2=\frac{1}{\,N\,}\,\left|\frac{M}{\sqrt{M}}\,\right|^2=\frac{M}{\,N\,}\;.\eeq
Equation (\ref{counting}) implies that the QFT can be used as an \textit{efficient} method for counting the number of terms in a uniform superposition --for it takes only $O(n^2)$ gates to implement it on a quantum computer \cite{Chuang}--, as long as $M/N$ and $1-M/N$ remain large enough, \ie do not decrease exponentially with the number of qubits, for this would require in practice an exponential precision in the estimation and therefore an exponential number of samples/repetitions of the protocol.

\section{The  Hardy-Littlewood-Ramanujan density matrices}
\label{HLR}

\noindent As explained in the main text,  the reduced density matrix of the Prime state with  $n$ qubits can be approximated by 
\beq
\bar{\rho}_A = \frac{1}{d} ( {\bf 1} + \ell_N {\bf C}_m ), \qquad \ell_N  \xrightarrow{N\rightarrow\infty}  \frac{1}{n \ln 2}  \, , 
\label{R1}
\eeq
where ${\bf C}_m$ is a matrix constructed with the even values of the Hardy-Littlewood constants  $C(2 k)$,
\beq
({\bf C}_m)_{i,j} = ( 1 - \delta_{i,j} )\, C( 2 |i-j|), \quad i, j =1, 2, \dots, d=2^{m-1}.
\label{R2}
\eeq
In order to justify the conjecture  \eqref{eigenvalues}
for the eigenvalues of ${\bf C}_m$, we define 
the density matrix 
\beq
{\rho}_D = \frac{1}{D} ( {\bf 1} + \ell_D  {\bf C}_D ), \qquad \ell_D   =  \frac{1}{2 \ln (2 D)}  \, , 
\label{R3}
\eeq
with 
\beq
({\bf C}_D)_{i,j} = ( 1 - \delta_{i,j} )\, C( 2 |i-j|), \quad i, j =1, 2, \dots, D.
\label{R4}
\eeq
If $D= 2^{m-1}$ and $n=2m$, Eq. \eqref{R3} and Eq. \eqref{R4} become Eq. \eqref{R1} and Eq. \eqref{R2} respectively. 
The Hardy-Littlewood  constants can be expressed as  \cite{GP99,GP06,Keating} 
\barray
C(h)    
& =  &     
\sum_{k=1}^\infty  \left( \frac{ \mu(k)}{ \phi(k)} \right)^2  c_k(h)  \, , 
 \label{R5}
\earray 
where
\barray
c_k(h)    
& =  &    \sum_{\substack{l=1 \\ {\rm gcd}(l,k)=1}} ^k 
  e^{2 \pi i \,h l/k}   \, , 
\label{R6} 
\earray 
are the Ramanujan's sums. Replacing  Eq. \eqref{R5} into Eq. \eqref{R4} gives 
\beq
({\bf C}_D)_{i,j} = ( 1 - \delta_{i,j} ) \sum_{k=1}^\infty  \left( \frac{ \mu(k)}{ \phi(k)} \right)^2  c_k(2|i-j|)),    \quad i, j =1, 2, \dots, D \, , 
\label{R7}
\eeq
which leads   us to write ${\bf C}_D$ as a sum of matrices, 
\beq
{\bf C}_D =  \sum_{k=1}^\infty  \left( \frac{ \mu(k)}{ \phi(k)} \right)^2  R_{k,D} \, ,   
\label{R8}
\eeq
where
\beq
({R}_{k,D})_{i,j} = ( 1 - \delta_{i,j} ) \,c_k( 2 |i-j|), \quad i, j =1, 2, \dots, D \, .
\label{R9}
\eeq
The sums $c_{k}(n)$ are multiplicative functions in $k$, namely 
\beq
c_{k_1 k_2}(h)  = c_{k_1}(h)\, c_{k_2}(h), \qquad {\rm gcd}(k_1, k_2)= 1 \, . 
\label{R10}
\eeq
In particular,
\beq
c_{2 k}(2 h)  = c_{2}(2 h)\, c_{k}(2h) = c_k(2h)\,, \qquad {\rm if} \; k \; {\rm odd}  \, , 
\label{R11}
\eeq
where we have used that $c_2(h) = \cos(\pi h)$, for integer $h$. Equation \eqref{R11}  implies that
\beq
{R}_{2 k,D} = { R}_{k,D},  \qquad {\rm if} \; k \;{\rm odd} \, . 
\label{R12}
\eeq
The M\"oebius and Euler's totient functions are also multiplicative, hence 
\beq
\mu(2 k ) = - \mu(k), \qquad \phi(2 k) = \phi(k),  \qquad {\rm if} \; k \;{\rm odd}  \, , 
\label{R13}
\eeq
where we have used that $\mu(2) = -1, \phi(2) = 1$. 
From the latter equations we  find  that  the sum 
 over the even integers $k$ in Eq. \eqref{R8} is equal to the sum over the odd integers, and thus 
\beq
{\bf C}_D =  2 \sum_{k=1 ({\rm odd} )}^\infty  \left( \frac{ \mu(k)}{ \phi(k)} \right)^2  R_{k,D}    \, . 
\label{R14}
\eeq
Our  aim is to find the eigenvalues of the matrix ${\bf C}_D$. This is 
a difficult problem in general. However,   we can find a good approximation performing  some simplifications. 
First of all,  we choose  $D$ as  the product of the first  $L$ odd prime numbers, that is
\beq
D =  p_2 \dots p_{L+1}  \, , 
\label{R15}
\eeq
where $p_2=3, p_3=5\dots$ .   Second, we truncate the sum in Eq. \eqref{R14} to the integers $k$ that divide $D$, 
obtaining the matrix
\beq
\tilde{{\bf C}}_D =  2 \sum_{k |D}   \left( \frac{ \mu(k)}{ \phi(k)} \right)^2  R_{k,D}    \, . 
\label{R16}
\eeq
We shall show below  that the  eigenvalues  of this matrix  provide a good 
approximation to those in Eq. \eqref{R14}. 
Using Eq.\eqref{R9} and Eq. \eqref{R6} one can write $R_{k,D}$ as
\beq
R_{k,D}  = \tilde{R}_{k, D} - \phi(k) \;   {\bf I}_D  \, , 
\label{R17}
\eeq
where  ${\bf I}_D$ is the identity matrix of dimension $D$,  and 
\beq
(\tilde{R}_{k, D})_{j, j'} = c_k( 2 (j - j') )  =   \sum_{\substack{l=1 \\ {\rm gcd}(l,k)=1}}^k   
  e^{4  \pi i\, l (j- j')/k}   \; .
\label{R18}
\eeq
The negative term in Eq. \eqref{R17}   guarantees that $(\tilde{R}_{k, D})_{j, j}=0$, because
\beq
 \sum_{\substack{l=1 \\ {\rm gcd}(l,k)=1}}^k    1 = \phi(k) \, . 
 \label{R19}
 \eeq
Replacing Eq. \eqref{R17} into Eq. \eqref{R16} gives 
\barray 
\tilde{{\bf C}}_D  & =   &   2 \sum_{k |D}    \left( \frac{ \mu(k)}{ \phi(k)} \right)^2  \tilde{R}_{k,D}  - 
2   \,  \sum_{k |D}    \frac{ \mu^2(k) }{ \phi(k) }  {\bf 1}_D 
\label{R20}     \\ 
& =   &   2 \sum_{k |D}    \left( \frac{ \mu(k)}{ \phi(k)} \right)^2  \tilde{R}_{k,D}  - 
2   \,  \frac{ D }{ \phi(D) }  {\bf 1}_D\,,
\nonumber 
\earray 
where we used the relation \cite{NumberTheory} 
\beq 
\sum_{d |n}    \frac{ \mu^2(d)  }{ \phi(d) } = \frac{n}{\phi(n)}  \, . 
 \label{R21}
 \eeq
 It turns out that the matrices  $\tilde{R}_{k,D}$ commute among themselves, so they can 
be diagonalized simultaneously and,  using Eq. \eqref{R20}, we can obtain the spectrum of  $\tilde{{\bf C}}_D$. 
This fact  follows from the following periodicity property of the  Ramanujan sums, \ie 
\beq
{\rm if} \; k | D \Longrightarrow c_k( 2( h + D) ) = c_k(2 h) \, , 
\label{R22}
\eeq
that can be easily proved using Eq. \eqref{R6}. Hence,  the eigenvalues,   $r_{k,D}(a) \;(a=1, \dots, D)$,   of $\tilde{R}_{k,D}$
are given by the Fourier transform of its entries,
\barray
r_{k,D}(a) &  =  &  \sum_{h=1}^D c_{k}( 2 h) e^{ - 2 \pi i h a/D}  = \sum_{\substack{l=1\\ {\rm gcd}(l,k)=1}}^k \sum_{h=1}^D 
\omega_{l, a}^h  , \qquad \omega_{l, a} = e^{ 2 \pi i\, \left( \frac{2 l}{k} - \frac{a}{D} \right)} \, . 
\label{R23}
\earray 
Let us first notice that   $\omega_{l, a}$  is a $D$ root of unity (use that $k|D$),
\beq
\omega_{l, a}^D =  e^{ 2 \pi i \left( \frac{2 l D}{k} - a \right)} = 1, \quad \forall l, \; \forall  a  \, . 
\label{R24}
\eeq
We have to consider two cases, when   $\omega_{l, a}=1$ and when   $\omega_{l, a}$ is a primitive root of unity. 
The former case occurs if   $a$ is given in terms of $l$ as
\beq
\omega_{l, a} = 1  \Longleftrightarrow a = \frac{ 2 l D}{k} \; ({\rm mod} \; D)  \Longrightarrow r_{k, D}(a)  = D  \, . 
\label{R25}
\eeq
Taking into account that ${\rm gcd}(l,k)=1$,  there are   $\phi(k)$ eigenvalues  of this type. 
In the case where  $\omega_{l, a}$ is  a primitive root of unity, one has  $\sum_{h}^D \omega_{l, a}^h =0$, and then 
\beq
\omega_{l, a} \neq  1   \Longleftrightarrow a \neq  \frac{ 2 l D}{k} \; ({\rm mod} \; D)  \Longrightarrow r_{k, D}(a)  = 0  \, . 
\label{R26}
\eeq
The number of vanishing eigenvalues is therefore  $D -  \phi(k)$. This proves   that the matrices 
$\tilde{R}_{k,D}$  are diagonal in the Fourier variables $a$.  This result, together with    Eq. \eqref{R20}, 
will allow us to obtain the eigenvalues of $\tilde{{\bf C}}_D$. Indeed, for a given  $a =1, \dots, D$,  the 
eigenvalue $r_{k,D}(a)$  can be either $D$ or $0$ depending on whether $a$  satisfies  condition  \eqref{R25} or \eqref{R26}. 
The latter conditions have to be analyzed for all values of $k|D$ and their contribution  added in 
Eq. \eqref{R20} with the corresponding weight $(\mu(k)/\phi(k))^2$. 

The previous  computation is simplified by the following remarkable fact: for each $a =1, \dots, D$,
there is only one pair $(l, k)$   where condition \eqref{R25} holds, while for the remaining   pairs  $(l', k')$ 
it is the  condition \eqref{R26} that is satisfied. The proof is as follows. Let $a$ be an integer  
satisfying \eqref{R25} for two pairs $(l_1, k_1)$  and $(l_2, k_2)$, that is, 
\beq 
a = \frac{ 2 l_1 D}{k_1}    \; ({\rm mod} \, D) = \frac{ 2 l_2 D}{k_2}   \; ({\rm mod} \, D) \, . 
\label{R27}
\eeq 
This equation  implies
\beq 
\left( \frac{ 2 l_1 }{k_1} -   \frac{ 2 l_2 }{k_2} \right)  D  = 0    \; ({\rm mod} \, D) \Longrightarrow 
2 \left( \frac{  l_1 }{k_1} -   \frac{  l_2 }{k_2} \right)  \in \Zmath  \, . 
\label{R28}
\eeq 
Since $1 \leq l_i  < k_i \; (i=1,2)$,  the term  $ \frac{  l_1 }{k_1} -   \frac{  l_2 }{k_2}$ belongs  to 
 the interval $(-1, 1)$.  Hence Eq. \eqref{R28} holds in three cases,
\beq 
2 \left( \frac{  l_1 }{k_1} -   \frac{  l_2 }{k_2} \right) = 0, \pm 1   \, 
\Longrightarrow 2 ( l_1 k_2 - l_2 k_1) = 0, \pm k_1 k_2  \, . 
\label{R29}
\eeq 
Recall that $k_i (i=1,2)$ are odd integers and so is $k_1 k_2$.  On the other hand, 
the l.h.s. of that equation is even, which shows that this condition cannot be fulfilled.
We are left with the case $l_1 k_2 = l_2 k_1$. Since $l_1$ does not divide $k_1$
there must exists an integer $x_1$ such that $l_1 =  l_2 x_1$. Similarly, 
$l_2 = l_1 x_2$ for some $x_2$. Using these equations we find
\beq
l_1 k_2 = l_2 k_1 \Longrightarrow x_1 k_2 = k_1\;  \& \;  k_2 = k_1 x_1 \Longrightarrow  x_1 x_2 = 1 
\Longrightarrow  x_1 = x_2  =1 \Longrightarrow  l_1 =  l_2 \; \& \; k_1 = k_2 \, ,
\label{R30}
\eeq
which means that the decomposition $a = 2 l D/k \; ({\rm mod} D)$ is unique, 
as stated above.  Table \ref{tab:appendix} shows the eigenvalues $r_{k,D}(a)$ for the case $D=15$. 
\begin{table} 
\centering
\begin{tabular}{|c|ccccccccccccccc|}
\hline
$k \setminus a$ & 1  & 2  & 3 & 4 & 5 & 6 & 7 & 8 & 9 & 10 & 11 & 12 & 13 & 14 & 15    \\
\hline 
$1$ & 0  & 0  & 0 & 0 & 0 & 0 & 0 & 0 & 0 & 0 & 0 & 0 & 0 & 0 & 1    \\
\hline
$3$ & 0  & 0  & 0 & 0 & 1 & 0 & 0 & 0 & 0 & 1 & 0 & 0 & 0 & 0 & 0    \\
\hline
$5$ & 0  & 0  & 1 & 0 & 0 & 1 & 0 & 0 & 1 & 0 & 0 & 1 & 0 & 0 & 0    \\
\hline
$15$ & 1  & 1  & 0 & 1 & 0 & 0 & 1 & 1 & 0 & 0 & 1 & 0 & 1 & 1 & 0    \\
\hline
\end{tabular}
\caption{The eigenvalues $r_{k,D}(a)/D$ for $D= 15$. The four rows corresponds
to the odd integers $k=1,3,5,15$ that divide $D$. 
The entry is 1 when $a$ is given 
by Eq. \eqref{R25}.}
\label{tab:appendix}
\end{table}

\begin{figure}[t!]
\centering
\includegraphics[width=0.55\textwidth]{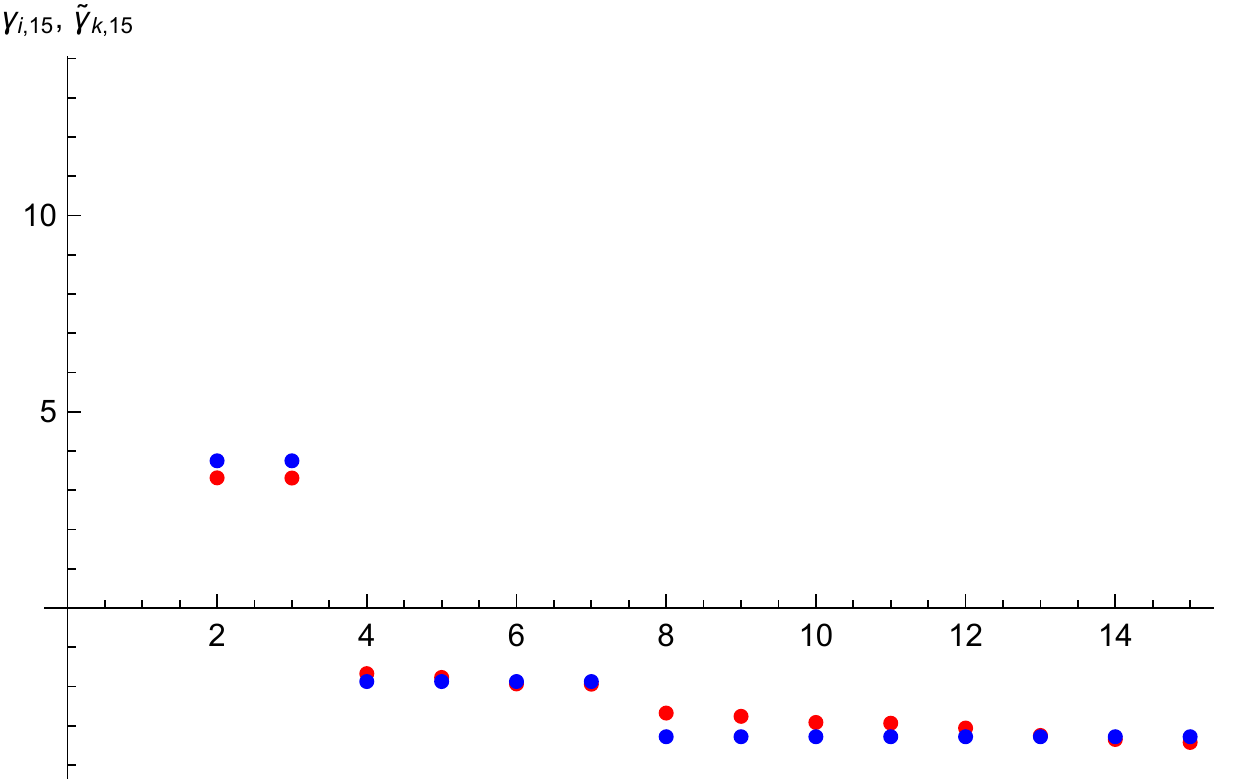}
\caption{Plot of the eigenvalues of the matrix ${\bf C}_D$: 
$\gamma_{i,D}$ 
for $i=2, \dots, 15$ (red dots),  and those of $\tilde{\bf C}_D$: 
 $\tilde{\gamma}_{k,D}$ (blue dots) for $D=15$ and $k=3,5,15$. The
highest eigenvalue (not plotted)  is given by $\gamma_{1, D} = 25.0742$ and $\tilde{\gamma}_{1,D} = 26.25$.}
\label{gamma15} 
\end{figure}

In summary,  the spectrum of the matrix $\tilde{R}_{k,D}$ is given by
\beq
{\rm Spec} \;  \tilde{R}_{k,D} = 
\left\{ \begin{array}{cc} 
D , & \phi(k) \\
0 , & D - \phi(k) \\
\end{array}  
\right.   \, . 
\label{R31}
\eeq
where the first column denotes the eigenvalue and the second one  its degeneracy. 
Using Eq. \eqref{R17} we obtain the spectrum of ${R}_{k,D}$, 
\beq
{\rm Spec} \;  {R}_{k,D} = 
\left\{ \begin{array}{cc} 
D  - \phi(k)  , & \phi(k) \\
- \phi(k) , & D - \phi(k) \\
\end{array}  
\right.   \, . 
\label{R32}
\eeq
which shows explicitly that it is a traceless matrix. 
Finally, using Eq. \eqref{R20}, together with the fact that  only one matrix $\tilde{R}_{k,D}$
contributes with $D$ to the total sum, we obtain the  eigenvalues of the matrix $\tilde{\bf C}_D$,
\beq
\tilde{\gamma}_{k, D} = 2 D   \left[    \frac{ 1}{ \phi^2(k)}   - \frac{1}{\phi(D)} \right] , \quad k |D 
\label{R33}
\eeq 
which has degeneracy $\phi(k)$.  We have simplified the notation using the fact that $\mu(k)^2 =1$ when 
$k|D$. 
Adding  these degeneracies over all the values of $k$ that divide $D$,  we recover the total
dimension of the matrix $\tilde{\bf C}_D$, using the relation \cite{NumberTheory} 
\beq
\sum_{k|D} \phi(k) = D \, . 
\label{R34}
\eeq
The matrix $\tilde{\bf C}_D$ was  introduced in order to obtain a simple analytic approximation
to the eigenvalues $\gamma_{i, D}$ of  the matrix ${\bf C}_D$. Fig. \ref{gamma15} shows the eigenvalues
of both  matrices for  $D=15$. Even for this small  value of $D$, they are numerically close and follow
the degeneracies given by $\phi(k)$, which in this case are $1,2,4,8$. 
Equation \eqref{R33} gives the positive and negative values of  $\tilde{\gamma}_{k, D}$.
The change of sign occurs at a value $k_D$ satisfying 
\beq
\phi^2(k_D) \simeq \phi(D)  \, . 
\label{R35}
\eeq

\subsection*{The Prime state}

\begin{figure}[b!]
\centering
\includegraphics[width=0.40\textwidth]{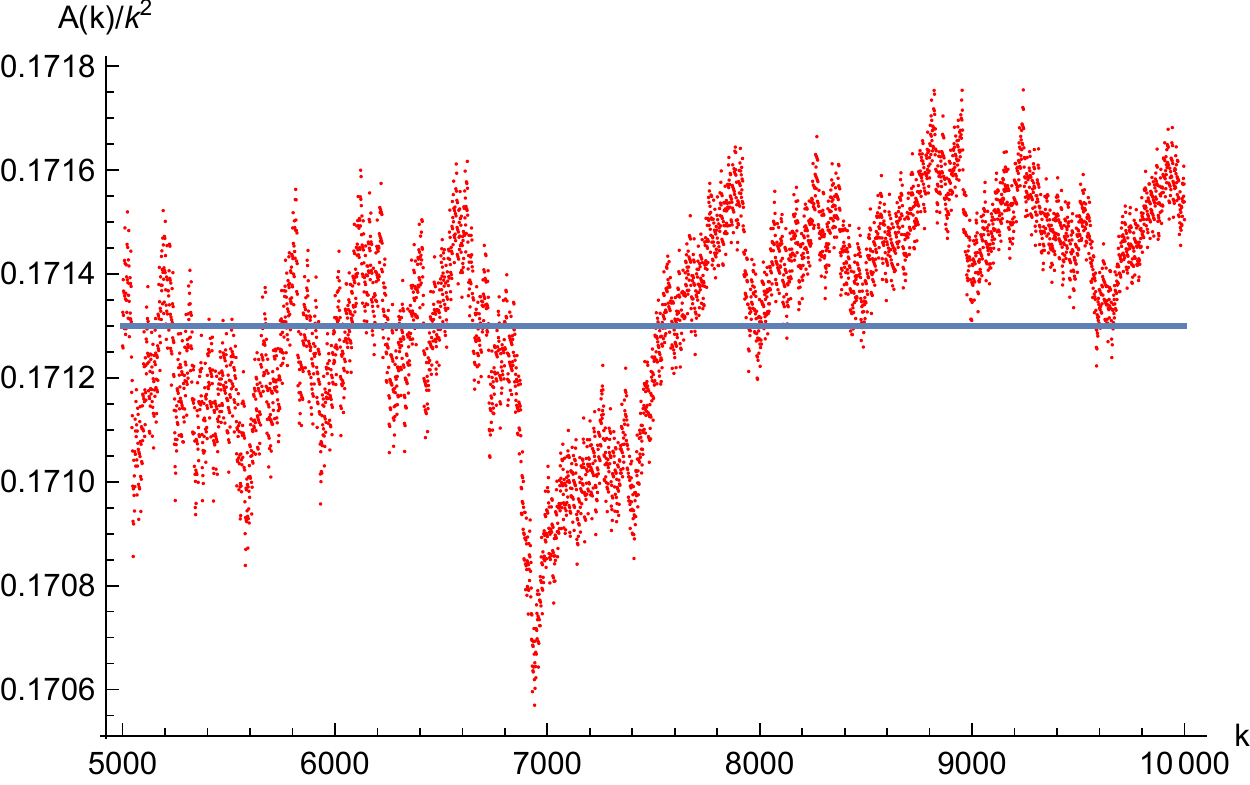}
\hspace{0.5cm}
\includegraphics[width=0.40\textwidth]{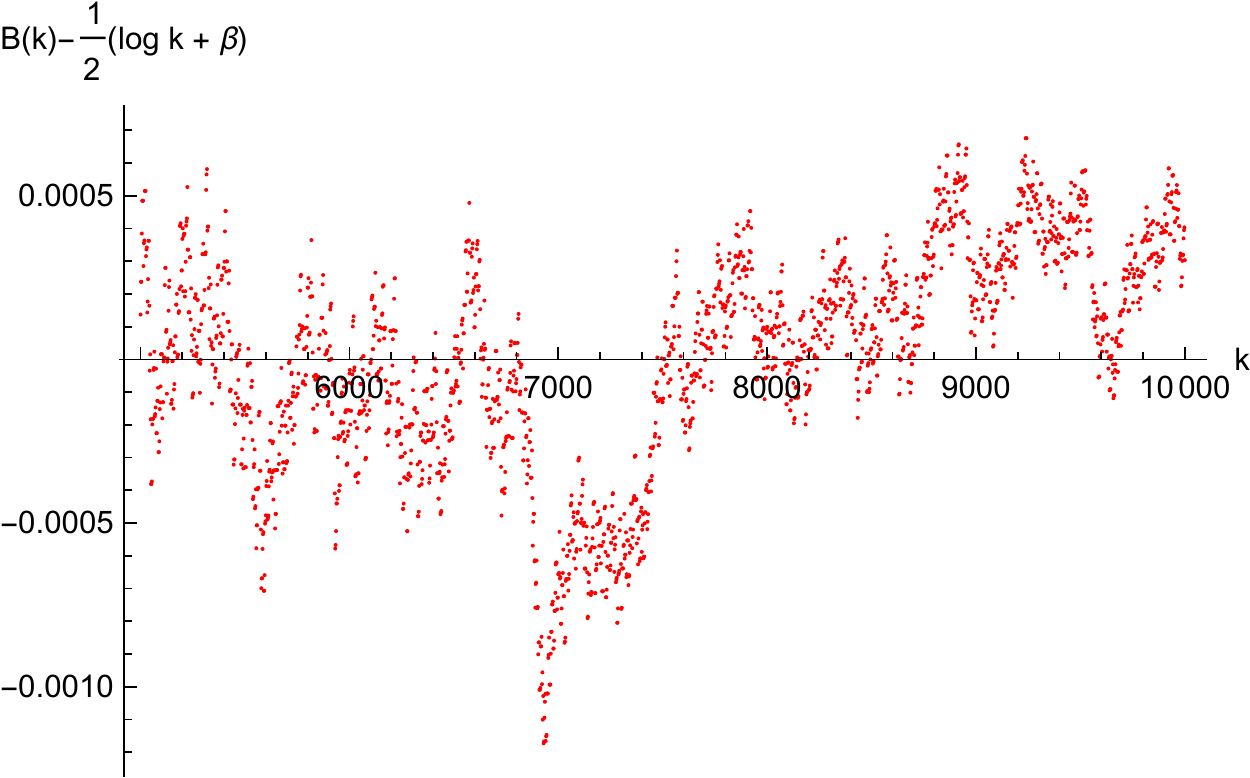}
\caption{Left: plot of $A(k)/k^2$. The straight line is $\alpha = 0.173304$. Right: plot of $B(k) - \frac{1}{2} ( \ln k + \beta)$ (see Eq. \eqref{R57}). 
Both plots are in the interval $k \in (5000, 10000)$.
}
\label{A-B} 
\end{figure}

\noindent Equation \eqref{R33} leads us to propose the approximation given in Eq. \eqref{eigenvalues} 
for  the eigenvalues of the matrix ${\bf C}_m$
\beq
\gamma_k \simeq 2^m \mu^2(k) \left( \frac{1}{ \phi^2(k)} - \frac{1}{\phi_m}  \right), \; k=1,3, \dots, k_m  \, , 
\label{R48}
\eeq
with a degeneracy $\phi(k)$.   The parameter  $k_m$ is   fixed imposing that 
the sum of the degeneracies is  the dimension of the matrix ${\bf C}_m$, that is,
\beq
A(k_m) \equiv  
\sum_{k =1 ({\rm odd})}^{k_m}  \mu^2(k) \phi(k) \simeq   2^{m-1}  \, , 
\label{R49}
\eeq 
where  $k_m$ is chosen as the highest square free  odd integer.  The asymptotic behaviour of the  sum is given by (see Fig. \ref{A-B}-left)
\beq
A(k) \simeq  \;  \alpha  k^2  + {\rm fluctuations} \quad (k \gg 1)  \, , 
\label{R50}
\eeq 
with 
\beq
\alpha = \frac{2}{5} \prod_{p} \left( 1 + \frac{ p-1}{p^2} \right) \left( 1 - \frac{1}{p} \right) = 0.171299873 \dots \label{R51}
\eeq
In this equation the product runs over all the prime numbers $\{p\}$. 
This result follows from the equation$^{[1]}$ 
 \footnotetext[1]{We thank A. C\'ordoba and F. Chamizo for providing us  this result. See also \cite{W27,L64}.}
\beq
\sum_{k =1}^{n}  \mu^2(k) \phi(k)  \simeq  \;  \;   \frac{1}{2} \prod_{p} \left( 1 + \frac{ p-1}{p^2} \right) \left( 1 - \frac{1}{p} \right) + {\rm fluctuations}
 \quad (n \gg 1)  \, . 
\label{R52}
\eeq 
Replacing Eq. \eqref{R50} into Eq. \eqref{R49} gives the asymptotic behaviour of $k_m$,
\beq
k_m \simeq   \left[ \frac{ 2^{m/2}}{ \sqrt{2 \alpha}}  \right]  \quad ( m \gg 1) \, , 
\label{R53}
\eeq
where $[x]$ denotes the integer part of $x$. 
To find $\phi_m$ we impose the vanishing of the trace of  ${\bf C}_m$, 
\beq
0 = 
{\rm tr} \, {\bf C}_m = \sum_{k =1 ({\rm odd})}^{k_m} \phi(k)  \gamma_k  \simeq  2^m 
\sum_{k =1 ({\rm odd})}^{k_m} \mu^2(k) \left( \frac{1}{ \phi(k)}  - \frac{\phi(k)}{\phi_m} \right)  \, ,
\label{R54}
\eeq
that yields
\beq
\phi_m = \frac{A(k_m)}{B(k_m)} ,
\label{R55}
\eeq
where 
\beq
B(k_m) \equiv  
\sum_{k =1 ({\rm odd})}^{k_m}  \frac{ \mu^2(k)}{  \phi(k)}   \, . 
\label{R56}
\eeq 
The asymptotic behaviour of this  sum is given by (see Fig. \ref{A-B}-right)
\beq 
B(k) \simeq \frac{1}{2} ( \ln k + \beta) + o(k^{- 1/2})\,, 
\label{R57}
\eeq
with
\beq
\beta = \gamma + \frac{1}{2} \ln 2 + \sum_p \frac{ \ln p}{ p(p-1)} = 1.679135304  \dots, 
\label{R58}
\eeq
where $\gamma$ is the Euler's constant. 
This result can be derived  from the equation$^{[1]}$ 
%
%
%
\beq
\sum_{k =1}^{n}  \frac{\mu^2(k)}{ \phi(k)}   \simeq  \;  \; \ln k + \gamma +  \sum_p \frac{ \ln p}{ p(p-1)} + o(n^{- 1/2}) 
 \quad (n \gg 1)  \, . 
\label{R59}
\eeq 
Using Eqs. \eqref{R49}, \eqref{R53} and \eqref{R57} we find
\beq
\phi_m \simeq  \frac{2^{m+1}}{m \ln 2 + \delta}, \qquad \delta = 2 \beta - \ln(2  \alpha) = 4.42946304048\dots
\label{R60}
\eeq

\begin{figure}[t!]
\centering
\includegraphics[width=0.37\textwidth]{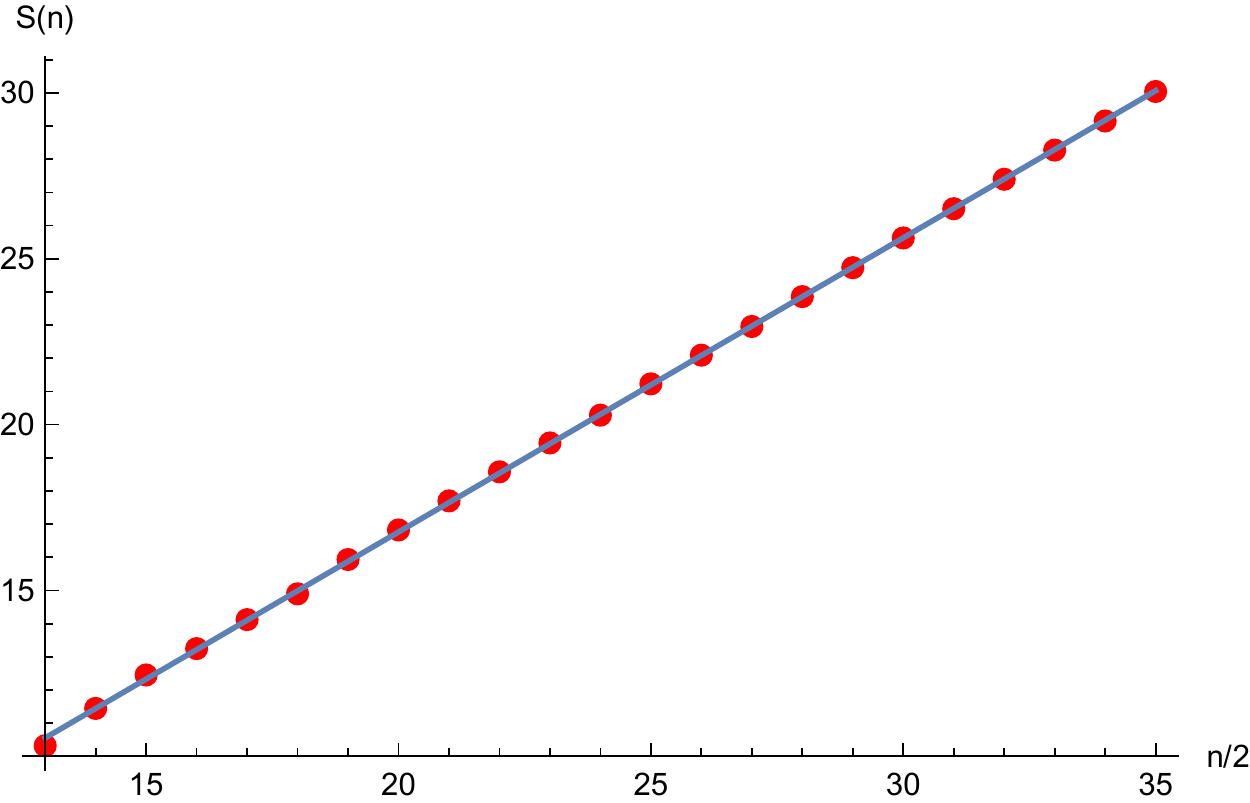}
\hspace{0.5cm}
\includegraphics[width=0.42\textwidth]{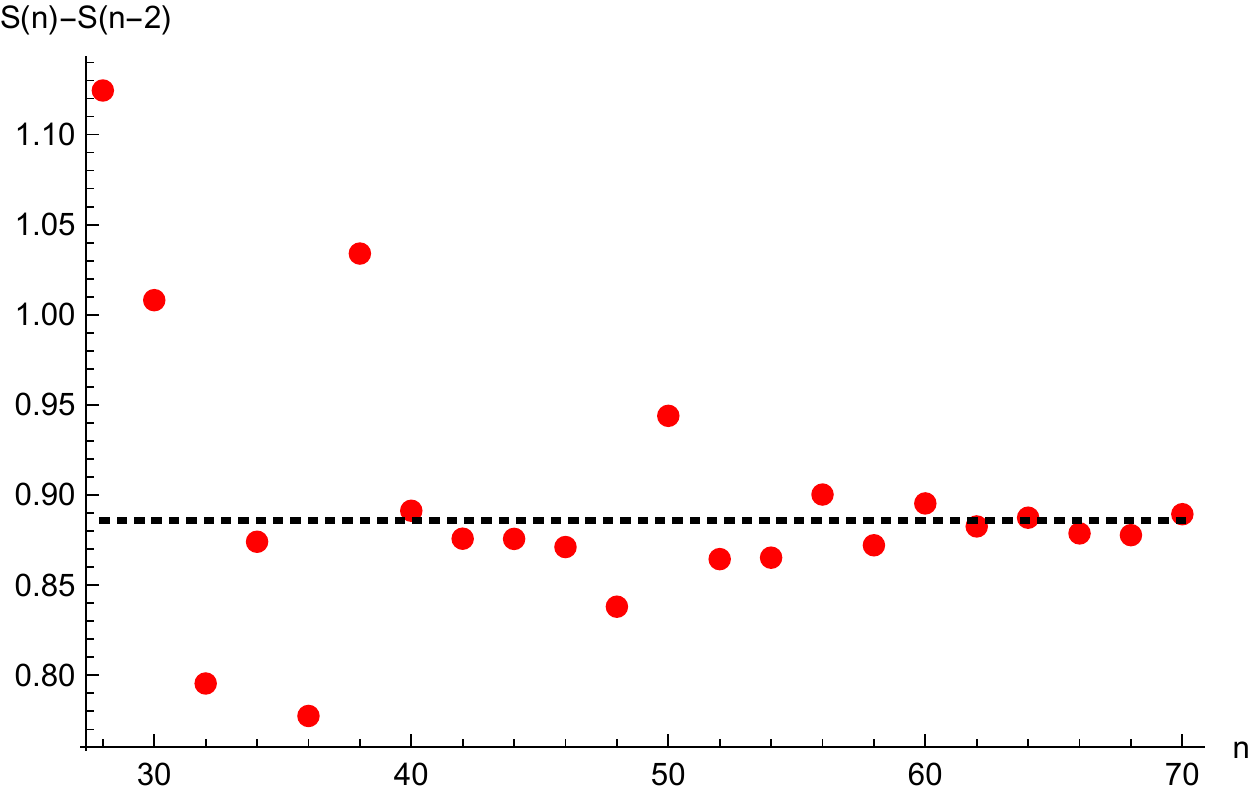}
\caption{Left: plot of $S(n)$ using Eq. \eqref{R62} in the interval $n/2 \in[13,35]$. The straight line is the linear fit
of the points, given by $0.886793 \frac{n}{2}  - 0.972872$. 
Righ: plot of $S(n)- S(n-2)$ in the interval $n \in[28,70]$. The dotted line is $H(3/\pi^2)$. 
}
\label{diffS} 
\end{figure}

The eigenvalues  of the  density matrix in Eq. \eqref{R1} can be approximated using Eq. \eqref{R48}
\barray
\lambda_k    & = &   \frac{1}{d} \,  |\mu(k)| ( 1 +  \ell_N  \gamma_k  ) , \qquad  \qquad k=1, 3, \dots, k_m  \, 
\label{R61} \\
& = & 2^{ 1-m} |\mu(k)|  \left( 1 + \frac{2^{m-1}}{m \ln 2 }  \left( \frac{1}{ \phi^2(k)} - \frac{1}{\phi_m}  \right) \right)  \,, 
\nonumber 
\earray 
that  yields the von Neumann entropy,
\barray
S(n)  & = & -  \sum_{k=1 ({\rm odd)}}^{k_m} \,   \phi(k) \,  \lambda_k \log_2 \lambda_k  \, ,
\label{R62}
\earray 
where $n = 2m$. Fig. \ref{diffS}-left shows the linear behaviour of $S(n)$ with a slope that seems to converge to the value
$H(3/\pi^2)$, as shown  in Fig. \ref{diffS}-right.

\section{Fourier expansion of the entropy of the Prime state for natural bi-partitions}

\noindent Given a natural bi-partition with the first (\ie least significant) $m$ qubits of the $n$-qubit Prime state,  $|\mathbb{P}_n\ra$, its von Neumann entropy is given by a function of the form
\beq \label{m,n} S(m,n) = n \,f\left(\frac{m}{n}\right)\,,\eeq
where $f\left(\frac{m}{n}\right)$ can be expressed as a Fourier series,
\beq \label{a,b} f\left(\frac{m}{n}\right) = \sum_{k=0}^\infty a_k\sin\left(2\pi k\, \frac{m}{n}\right)+ b_k\cos\left(2\pi k \,\frac{m}{n}\right)\,.\eeq
Alternatively, Eq. \eqref{m,n} can be normalized by the entropy of the half chain, $S(\frac{n}{2})$, as in Eq. \eqref{m/n}. This results in a Fourier series analogous to that in Eq. \eqref{a,b}, but with the coefficients divided by $f(\frac{1}{2})=0.4430\dots$ Such latter coefficients, $\{a'_k,b'_k\}$, obtained numerically for $k=0,1,\dots,7$, are presented in Table.\ref{tab:fourier}. The resulting truncated Fourier series, compared to the exact values appearing in Fig. \ref{fig:M}, is shown in Fig. \ref{fig:fourier}.

\begin{table}[h!] 
    \centering
	\begin{tabular}{|c|c|c|} \hline
		
		$k$ & $a'_k$ & $b'_k$ \\\hline
		0 &   0    & 5.38987832e-01 \\\hline
		1 & -6.41605273e-02 & -4.46006651e-01 \\\hline
		2 & -8.36285927e-03 & -2.72896597e-02 \\\hline
		3 & -1.80403548e-02 & -3.91333186e-02 \\\hline
		4 & -4.62380210e-03 & -6.47395477e-03 \\\hline
		5 & -8.37937907e-03 & -9.22366149e-03 \\\hline
		6 & -3.03787769e-03 & -2.01306807e-03 \\\hline
		7 & -4.11451014e-03 & -2.29208566e-03 \\\hline

	\end{tabular}
	\caption{Coefficients of the Fourier expansion of the entropy of the Prime state $|\mathbb{P}_n\ra$, normalized by the entropy of half chain, for natural bi-partitions of size $m$, and for $k=0,1,\dots,7$.}
	\label{tab:fourier}
\end{table}

\begin{figure}[h!]
	\centering
	\includegraphics[scale=0.6]{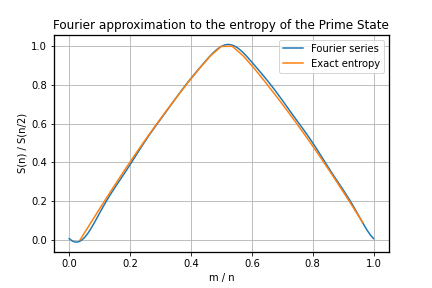}
	
	\caption{\small{Comparison of the truncated Fourier series obtained with the coefficients given in Table \ref{tab:fourier}}, and the exact entropies plotted in Fig. \ref{fig:M}.}
	\label{fig:fourier}
\end{figure}

\section{Reduced density matrix of the odd composite state}

\noindent We provide here the derivation of the asymptotic expression for the reduced density matrix of the odd composite state upon tracing out the last (\ie most significant) $n-m$ qubits. We begin with the diagonal elements,
\beq \label{diagonal} \rho_{ii} = \frac{2^{n-m}-\pi_{M,i}(N)}{N/2-\pi(N)}\quad,\quad i \;\;odd \,,\eeq
where $N=2^n$ and $M=2^m$. The off-diagonal elements are given by
\beq \label{off-diagonal} \rho_{ij} = \frac{2^{n-m}-\pi_{M;\, i,j}(N)}{N/2-\pi(N)}\quad,\quad i,j \;\;odd\,. \eeq
In the limit $n,m\rightarrow\infty$, Eq. \eqref{diagonal} and Eq. \eqref{off-diagonal} become
\beq \rho_{ii} = \frac{2^{n-m}-N/(\phi(M)\ln\,N)}{N/2-N/\ln\,N}\quad,\quad i \;\;odd\  \,,\eeq
and
\beq  \rho_{ij} =  \frac{2^{n-m}-NC(|i-j|)/(\phi(M)\ln^2\,N)}{N/2-N/\ln\,N}\quad,\quad i,j \;\;odd \,,\eeq
respectively. Substituting $\phi(M)=2^{m-1}$ and rearranging terms, we obtain
\beq \label{Diagonal} \rho_{ii} = \frac{2^{-m}\,(1-2/(n\ln\,2))}{1/2-1/(n\ln\,2)}\quad,\quad i \;\;odd\  \,,\eeq
and
\beq \label{Off-diagonal} \rho_{ij} =  \frac{2^{-m}(1-2C(|i-j|)/(n^2\ln^2\,2))}{1/2-1/(n\ln\,2)}\quad,\quad i,j \;\;odd \,.\eeq
Combining Eq. \eqref{Diagonal} and Eq. \eqref{Off-diagonal}, we obtain, in the limit $n\rightarrow\infty$, the expression in the main text,
\beq  \overline{\rho}_A = \frac{1}{d} \left(\mathbbm{1}+{\bf P}_m \right)\,.\eeq

\end{appendices}

\begin{thebibliography}{99}
\bibitem{natural} We consider natural bi-partitions those that separate the first $m$ qubits and the remainder $n-m$ qubits, without any reordering.
\bibitem{errors}  If $p_i$ is the exact probability, and $\hat{p}_i$ the estimate of it, the statistical additive error is defined as $p_i=\hat{p}_i+\epsilon_i$, and the multiplicative error as $\hat{p}_i=(1+\varepsilon_i)\,p_i$.
\end{thebibliography}
\end{document}